%% file: text14-clean.tex
\documentclass[useAMS,usenatbib]{mn2e}
\usepackage{graphicx}
\usepackage{color}
\usepackage{ulem}
\def\apj{ApJ}
\def\apjs{ApJS}

\def\aap{A\&A}
\def\aaps{A\&AS}
\def\aj{AJ}
\def\mnras{MNRAS}

\def\pasp{PASP}

\def\araa{Ann.Rev.Astron.Astrophys.}

\def\rmxaa{Revista Mexicana de Astronomia y Astrofisica}

\title[J0811$+$4730: the most-metal poor galaxy]
{J0811$+$4730: the most metal-poor star-forming dwarf galaxy known}
\author[Y. I. Izotov et al.]{Y. I.\ Izotov$^{1}$,
T. X.\ Thuan$^{2}$, N. G.\ Guseva$^{1}$ and S. E.\ Liss$^{2}$\\
                $^{1}$Main Astronomical Observatory,
                     Ukrainian National Academy of Sciences,
                     Zabolotnoho 27, Kyiv 03143,  Ukraine,\\
                     izotov@mao.kiev.ua, guseva@mao.kiev.ua\\
                $^{2}$Astronomy Department, University of Virginia, 
                     P.O. Box 400325, Charlottesville, VA 22904-4325,\\
                     txt@virginia.edu, sel7pa@virginia.edu\\
}
\begin{document}


\pagerange{\pageref{firstpage}--\pageref{lastpage}} \pubyear{2012}

\maketitle

\label{firstpage}

\begin{abstract}
We report the discovery of the most metal-poor dwarf star-forming galaxy (SFG)
known to date, J0811$+$4730. 
This galaxy, at a redshift $z$=0.04444, has a Sloan Digital Sky
Survey (SDSS) $g$-band absolute magnitude $M_g$ = $-$15.41 mag. It
was selected by inspecting the spectroscopic data base in the Data 
Release 13 (DR13) of the SDSS. 
LBT/MODS spectroscopic observations reveal its oxygen abundance to be 
12 + log O/H = 6.98 $\pm$ 0.02, the lowest ever observed for a SFG. 
J0811$+$4730 strongly deviates from the main-sequence defined by 
SFGs in the emission-line diagnostic diagrams and the 
metallicity -- luminosity diagram. These differences are caused mainly
by the extremely low oxygen abundance in J0811$+$4730, which is $\sim$ 10 times
lower than that in main-sequence SFGs with similar luminosities. 
By fitting the spectral energy distributions of the 
SDSS and LBT spectra, we derive a stellar mass of $M_\star$ = 
10$^{6.24} - 10^{6.29}$ M$_\odot$
(statistical uncertainties only), and we find that a considerable fraction of
the galaxy stellar mass was formed during the most recent burst of star 
formation.
\end{abstract}


\section{Introduction}\label{sec:INT}

Extremely metal-deficient 
star-forming galaxies (SFGs) with oxygen abundances
12+log~O/H$\la$7.3 constitute a rare but important class of galaxies
in the nearby Universe. They are thought to be the best local analogs of the
numerous population of the dwarf galaxies at high 
redshifts that played an important role in the reionization of the Universe 
at redshifts $z$ $\sim$ 5--10 \citep[e.g. ][]{O09,K17}. 
Their proximity permits studies of their stellar,
gas and dust content with a sensitivity and spectral resolution that are not 
possible for high-$z$ galaxies. 

Many efforts have been made in the past to discover the most metal-poor nearby
galaxies. The galaxy I~Zw~18, first spectroscopically observed by \citet{SS72},
with oxygen abundances 12+log~O/H~$\sim$~7.17--7.26 in its two brightest
regions \citep[e.g. ][]{SK93,IT98} stood as the lowest-metallicity SFG for a 
long period. It was replaced by SBS~0335$-$052W with an  
oxygen abundance 12+log~O/H=7.12$\pm$0.03, 
derived from the emission of the entire galaxy \citep{I05},
and oxygen abundances in its two brightest knots of star formation 
of 7.22 and 7.01 \citep{I09}. 

The large data base of the Sloan Digital Sky Survey (SDSS) offered a possibility
for a systematic search of the most metal-deficient SFGs. 
In particular, \citet{I12} and \citet{G17}
found two dozens of galaxies with 12+log~O/H$<$7.35 in the 
SDSS. However, no SFG with a metallicity below that
of SBS 0335$-$052W was found.

Local galaxies with very low-metallicities have also been discovered in other 
contexts. 
\citet*{P05} 
studying the properties of dwarf galaxies in the Cancer-Lynx void
found that the weighted mean oxygen abundance in five regions of 
the irregular dwarf galaxy DDO~68 is 12+log~O/H=7.21$\pm$0.03.

Two extremely metal-poor dwarf galaxies were discovered in the
course of the  Arecibo Legacy Fast ALFA survey \citep[ALFALFA, ][]{G05,H11}.
\citet{S13} found that the oxygen abundance in the nearby dwarf galaxy 
AGC~208583 = Leo~P is 7.17$\pm$0.04, still more metal-rich than SBS~0335$-$052W.
Finally, \citet{H16} showed that the nearby dwarf galaxy AGC~198691
has an oxygen abundance of 12~+~logO/H~=~7.02$\pm$0.03, lower than the 
average value in SBS~0335$-$052W.

\input{tab1_1.tex}

In this paper, we present Large Binocular Telescope (LBT)\footnote{The LBT 
is an international collaboration among institutions in the United States, 
Italy and Germany. LBT Corporation partners are: The University of Arizona on 
behalf of the Arizona university system; Istituto Nazionale di Astrofisica, 
Italy; LBT Beteiligungsgesellschaft, Germany, representing the Max-Planck 
Society, 
the Astrophysical Institute Potsdam, and Heidelberg University; The Ohio State 
University, and The Research Corporation, on behalf of The University of Notre 
Dame, University of Minnesota and University of Virginia.} spectroscopic
observations with high signal-to-noise ratio of the compact dwarf SFG 
J0811$+$4730. This galaxy stood out by its emission-line ratios, 
during the inspection of the  
spectra in the SDSS Data Release 13 (DR13) data base \citep{A16}, as 
potentially having a very low metallicity. 
Its coordinates, redshift and other characteristics obtained from the SDSS
photometric and spectroscopic data bases are shown in Table \ref{tab1}.

The LBT observations and data reduction are described in 
Sect.~\ref{sec:observations}. We derive element abundances in 
Sect.~\ref{sec:abundances}. Integrated characteristics of J0811$+$4730 are
discussed in Sect.~\ref{sec:integr}. Emission-line diagnostic diagrams and
the metallicity-luminosity relation for a sample of the most-metal poor
SFGs, including J0811$+$4730, are considered in Sect.~\ref{sec:diagrams}.
Finally, in Sect.~\ref{sec:conclusions} we summarize our main results.

\begin{figure*}
\centering{
\includegraphics[angle=-90,width=0.90\linewidth]{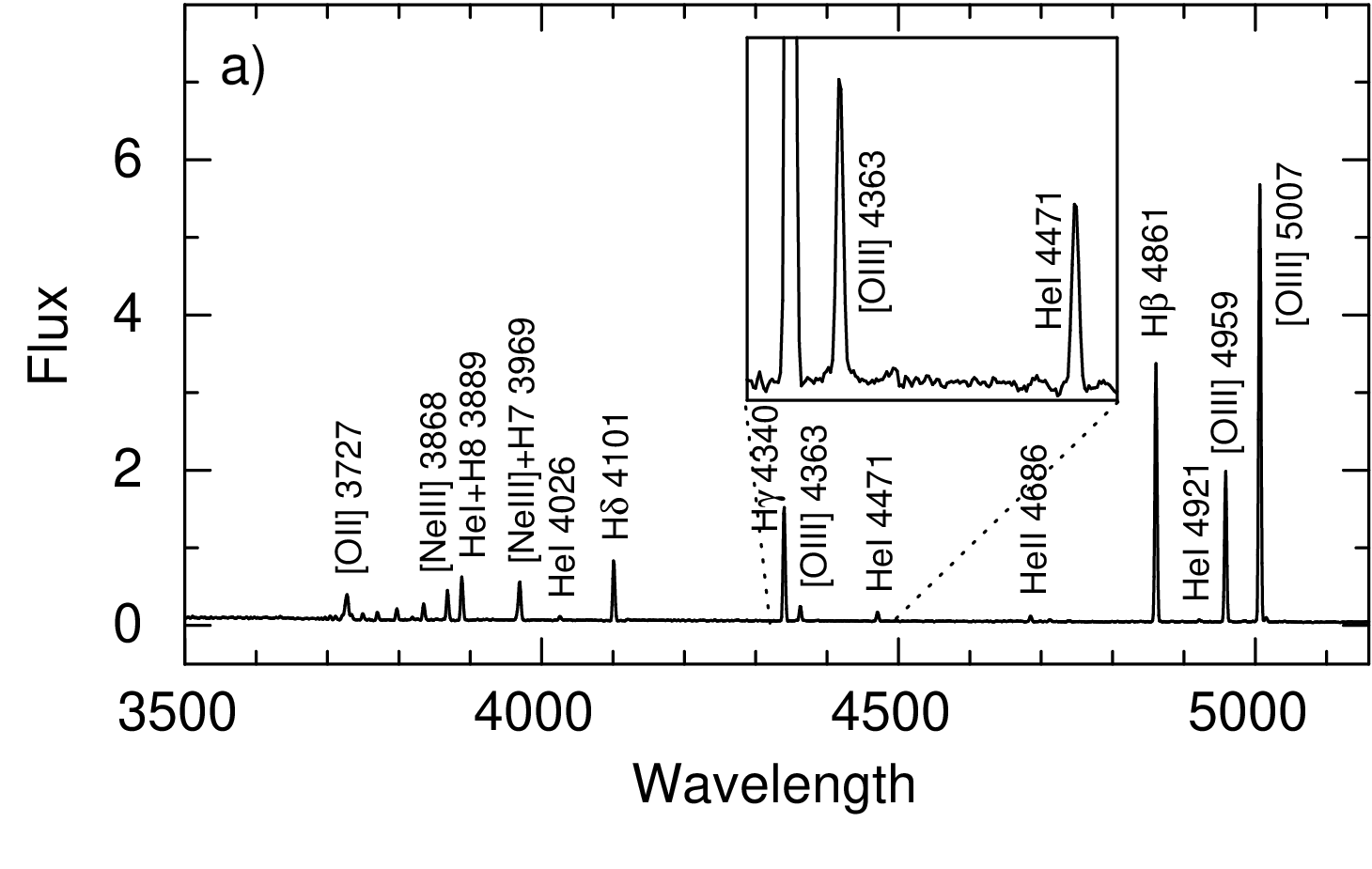}
\includegraphics[angle=-90,width=0.90\linewidth]{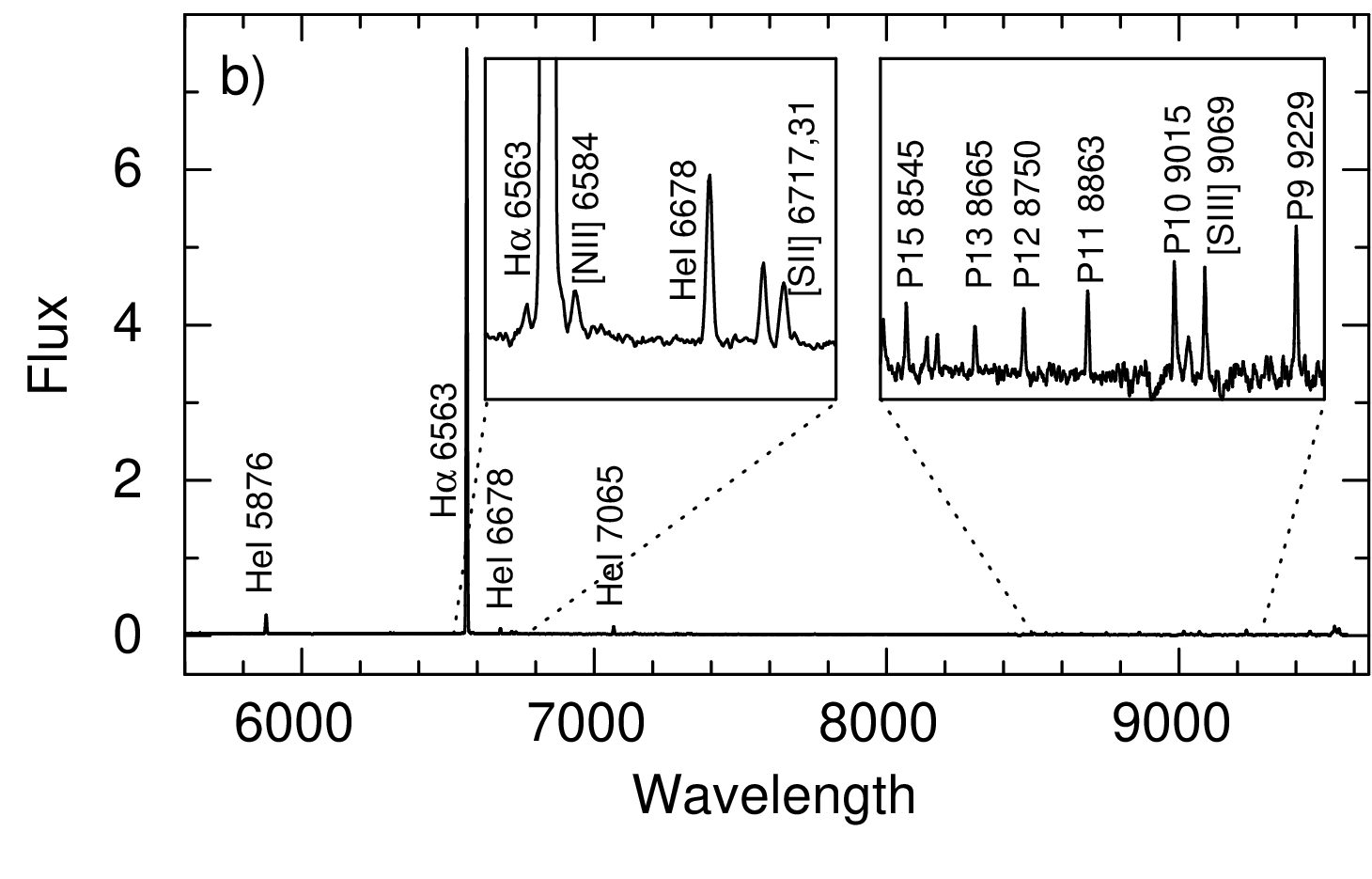}}
\caption{The rest-frame LBT spectrum of J0811$+$4730 uncorrected for extinction.
Insets show expanded parts of spectral regions in the rest 
wavelength ranges 4320\AA\ -- 4490\AA, 6520\AA\ -- 6770\AA\ and 
8500\AA\ -- 9280\AA\ for a better
view of weak emission lines. Some interesting emission lines are 
labelled. Wavelengths are in \AA\ and fluxes are in 
units of 10$^{-16}$ erg s$^{-1}$ cm$^{-2}$ \AA$^{-1}$.}
\label{fig1}
\end{figure*}

\section{LBT Observations and data reduction}\label{sec:observations}

We have obtained LBT long-slit spectrophotometric observations of J0811$+$4730 
on 1 February, 2017 in the twin binocular mode using both the MODS1 and 
MODS2 spectrographs\footnote{This paper used data obtained with the MODS 
spectrographs built with
funding from NSF grant AST-9987045 and the NSF Telescope System
Instrumentation Program (TSIP), with additional funds from the Ohio
Board of Regents and the Ohio State University Office of Research.}, 
equipped with two 8022 pix $\times$ 3088 pix CCDs. The G400L grating in 
the blue beam, with a dispersion of 0.5\AA/pix, and the G670L grating in the 
red beam, with a dispersion of 0.8\AA/pix, were used. Spectra were obtained in 
the wavelength range 3200 -- 10000\AA\ with a 1.2 arcsec wide slit, 
resulting in a resolving power $R$ $\sim$ 2000. The seeing during 
the observations was in the range 0.5 -- 0.6 arcsec.

Six 890~s subexposures were obtained with each spectrograph, resulting
in the total exposure of 2$\times$5340~s counting both spectrographs.
The airmass varied from 1.16 for the first subexposure to 1.05 for
the sixth subexposure. 
The slit was oriented at the fixed position angle of 80${\degr}$. This is to be 
compared with the parallactic angles of 74${\degr}$ during the first subexposure
and of 53${\degr}$ during the sixth subexposure. According to \citet{F82}, such
differences between the position angle and the parallactic angle would result in
an offset perpendicular to the slit of the [O~{\sc ii}] $\lambda$3727 
wavelength region relative to the H$\beta$ wavelength region of less than 
0.1 -- 0.2 arcsec. Thus, 
the effect of atmospheric refraction is small for all subexposures. 

The spectrum of the spectrophotometric standard star GD~71, obtained during 
the same night with a 5 arcsec wide slit, was used for flux calibration.
It was also used to correct the red part of the 
J0811$+$4730 spectrum for telluric absorption lines. 
Additionally, calibration frames of biases, flats and comparison lamps 
were obtained during the daytime, after the observations.

The MODS Basic CCD Reduction package {\sc modsCCDRed}\footnote{http://www.astronomy.ohio-state.edu/MODS/Manuals/ MODSCCDRed.pdf} was used for bias subtraction 
and flat field correction, while wavelength and flux calibration was 
done with {\sc iraf}\footnote{{\sc iraf} is distributed by the 
National Optical Astronomy Observatories, which are operated by the Association
of Universities for Research in Astronomy, Inc., under cooperative agreement 
with the National Science Foundation.}. 
Finally, all MODS1 and MODS2 subexposures were 
combined. The one-dimensional spectrum 
of J0811$+$4730 shown in Fig. \ref{fig1} was extracted 
in a 1.2 arcsec aperture along the spatial axis. Strong emission lines are 
present in this spectrum, suggesting active star formation.
In particular, a strong [O~{\sc iii}] $\lambda$4363 emission line is 
detected, allowing a reliable abundance determination. Because of the
importance of this line, we have checked whether its flux can be affected by 
such events as cosmic ray hits. We find that the flux ratio of 
[O~{\sc iii}]$\lambda$4363 to the nearest H$\gamma$ $\lambda$4340 emission line 
is nearly constant in all MODS1 and MODS2 subexposures, varying by not more than
1 -- 2 per cent. This implies that the [O~{\sc iii}]$\lambda$4363 emission line 
is not affected by cosmic ray hits.

Emission-line fluxes were measured using the  {\sc iraf splot} routine. 
The errors of the line fluxes were calculated from the
photon statistics in the non-flux-calibrated spectra, and by adding a 
relative error of 1 per cent in the 
absolute flux distribution of the spectrophotometric standard star.
They were then propagated in the calculations of the elemental 
abundance errors. 

\input{tab2.tex}

\input{tab3_2.tex}

The observed fluxes were corrected for extinction, 
derived from the observed decrement of the hydrogen Balmer emission lines
H$\alpha$, H$\beta$, H$\gamma$, H$\delta$, H9, H10, H11, and H12.
Two lines, H7 and H8, were however excluded because they are blended with
other lines.
The equivalent widths of the underlying stellar Balmer absorption lines, 
assumed to be the same for each line, were derived simultaneously with the 
extinction in an iterative manner, following \citet{ITL94}. They are shown in 
Table \ref{tab2}. All hydrogen lines were corrected for underlying
absorption, in addition to correction for extinction. It is seen from 
Table \ref{tab2} that the corrected fluxes of the H9 -- H16 emission lines
are somewhat larger than the case B values. This is probably due to an 
overcorrection of underlying absorption for these higher order lines. However, 
this overcorrection makes
little difference on the determination of extinction which depends mainly on the
observed decrement of the H$\alpha$, H$\beta$, H$\gamma$ and H$\delta$ emission
lines.

We note that excluding H$\alpha$, the only Balmer 
hydrogen line observed in the red part of the spectrum, in the extinction 
determination would result in a somewhat higher extinction coefficient,  
$C$(H$\beta$) = 0.185 compared to the value of 0.165 in Table \ref{tab2}.
This disagreement is small, within the errors. It probably comes from a slight 
mismatch between the blue and red parts of the spectrum, not exceeding a few
per cent of the continuum flux in the overlapping wavelength range. 
We have adopted the value of 0.165. It is somewhat higher than the $C$(H$\beta$)
expected for extremely low-metallicity SFGs. However, part of this 
extinction is due to the Milky Way with relatively high
$A(V)_{\rm MW}$ = 0.180 mag according to 
the NASA/IPAC Extragalactic Database (NED) and corresponding to 
$C$(H$\beta$)$_{\rm MW}$=0.085. Therefore, the 
internal extinction coefficient in J0811$+$4730 is only
$C$(H$\beta$)$_{\rm int}$=0.080. We have adopted the reddening law by 
\citet*{C89} with a total-to-selective extinction ratio $R(V)$ = 3.1.

The observed emission-line fluxes $F(\lambda)$/$F$(H$\beta$) multiplied by 100, 
the emission-line fluxes corrected for underlying absorption and extinction $I(\lambda)$/$I$(H$\beta$) multiplied by 100, the extinction 
coefficient $C$(H$\beta$), the rest-frame equivalent width EW(H$\beta$),
the equivalent width of the Balmer absorption lines, and the observed
H$\beta$ flux $F$(H$\beta$) are listed in Table \ref{tab2}. We note that 
the EW(H$\beta$) in J0811$+$4730 is high, $\sim$~280\AA, implying that 
its optical emission is dominated by radiation from a very young burst, with
age $\sim$ 3 Myr.

It is seen in Fig. \ref{fig1} that the ratio of H$\alpha$ and H$\beta$
peak intensities (slightly larger than 2) is lower than the ratio of their 
total fluxes (about a factor of 3). This is in large
part due to the somewhat different spectral resolution of the blue and red 
spectra of J0811$+$4730, as they were obtained with different gratings.
Indeed, the full widths at half maximum (FWHMs) of H$\alpha$ and H$\beta$ are 
4.4\AA\ and 2.9\AA, respectively. These widths are larger than the widths of 
the Ar lines of 3.6\AA\ and 2.2\AA\ in the blue and red comparison spectra,
respectively, indicating that both profiles are 
partially resolved. By deconvolving the instrumental profiles from the  
observed profiles, we obtain an intrinsic 
velocity dispersion $\sigma$ of $\sim$45$\pm$5 km s$^{-1}$ for both the 
H$\alpha$ and H$\beta$ emission lines. This value is separately derived from each of the MODS1 and MODS2 spectra. The uncertainty is derived from the Gaussian fitting of both lines, followed by
the deconvolution of their profiles.The velocity dispersion value is fully consistent with the one expected from the $L$(H$\beta$) --
$\sigma$ relation for supergiant H~{\sc ii} regions \citep[e.g. ][]{C12}, 
and supports the assumption that the emission lines originate from a single 
H~{\sc ii} region. 

\section{Element abundances}\label{sec:abundances}

The procedures described by \citet{I06} are used to determine element 
abundances from the LBT spectrum. We note that SDSS spectra of
J0811$+$4730 are available in Data Releases 13 and 14, but they cannot be used for  
abundance determination because the strong lines are clipped in those spectra. 
The temperature $T_{\rm e}$(O~{\sc iii}) is calculated from the 
[O~{\sc iii}] $\lambda$4363/($\lambda$4959 + $\lambda$5007) emission-line flux 
ratio. It is used to derive the abundances of O$^{2+}$, Ne$^{2+}$ and Ar$^{3+}$.
To obtain the abundances of O$^{+}$, N$^{+}$, S$^{+}$ and Fe$^{2+}$, the electron
temperature $T_{\rm e}$(O~{\sc ii}) needs to be derived. In principle, this can
be done by using the [O~{\sc ii}]$\lambda$3727/($\lambda$7320+$\lambda$7330) flux
ratio. However, the latter two lines are extremely weak in the spectrum 
of J0811$+$4730,
making such a determination of $T_{\rm e}$(O~{\sc ii}) not possible. Similarly,
the electron temperature $T_{\rm e}$(S~{\sc iii}) is needed to derive the 
S$^{2+}$ and Ar$^{2+}$ abundances. It can be obtained from the 
[S~{\sc iii}]~$\lambda$6312/($\lambda$9069+$\lambda$9531) flux ratio. However,
the [S~{\sc iii}]~$\lambda$6312 line is very weak (Table \ref{tab2}) and
the [S~{\sc iii}]~$\lambda$9069 line, at the J0811$+$4730 redshift, falls at the wavelength of $\sim$9475\AA, in a region with numerous night sky emission lines
\citep[e.g., fig. 9 in ][]{L11} and telluric absorption lines
\citep[e.g., fig. 1 in ][]{R16}. Furthermore, the strongest 
[S~{\sc iii}]~$\lambda$9531 line is at the very edge of the LBT spectrum, where
the sensitivity is low. 
All these factors make the determination of 
$T_{\rm e}$(S~{\sc iii}) somewhat uncertain. In fact, using the 
[S~{\sc iii}]$\lambda$6312 and [S~{\sc iii}]$\lambda$9069 fluxes from 
Table \ref{tab2}, equations from \citet{A84} and adopting 
[S~{\sc iii}]$\lambda$9531/$\lambda$9069 = 2.4 we derive the high and very 
uncertain value $T_{\rm e}$(S~{\sc iii}) = 27000$\pm$6000K.

Therefore we have used the expressions of \citet{I06}, obtained from 
photoionized H~{\sc ii} region models, to derive the electron 
temperatures $T_{\rm e}$(O~{\sc ii}) and $T_{\rm e}$(S~{\sc iii}). We have also
adopted the errors for these temperatures to be equal to the error for
the electron temperature $T_{\rm e}$(O~{\sc iii}).
The errors quoted  for $T_{\rm e}$(O~{\sc ii}) and $T_{\rm e}$(S~{\sc iii}) should 
be considered as lower limits as they do not take into account the uncertainty 
introduced by our reliance on 
grids of photoionization models to determine these temperatures. These models 
show inevitably a 
dispersion due to varying parameters such as ionization parameter, electron 
number density, chemical composition, etc., that will increase the uncertainty 
in these two temperatures. 
 The electron number density $N_{\rm e}$(S~{\sc ii}) is derived from the 
[S~{\sc ii}] $\lambda$6717/$\lambda$6731 emission line ratio. 
However, it is not possible to obtain the electron number density
from the [O~{\sc ii}]$\lambda$3726/$\lambda$3729 flux ratio because of the
insufficient spectral resolution. 

The total oxygen abundance is derived as follows:
\begin{equation}
\frac{\rm O}{\rm H} = \frac{{\rm O}^++{\rm O}^{2+}+{\rm O}^{3+}}{{\rm H}^+}, 
\label{OH} 
\end{equation}
where the abundances of ions O$^+$, O$^{2+}$, O$^{3+}$ are obtained using 
the relations of \citet{I06}. For ions of other heavy elements, we also use
the relations of \citet{I06} to derive the ionic abundances, the ionization
correction factors and the total heavy element abundances.

The electron temperatures, electron number densities, ionic abundances,
ionization correction factors and total heavy element abundances are presented 
in Table~\ref{tab3}. The electron temperature $T_{\rm e}$(O~{\sc iii})
= 21700 $\pm$ 500 K is derived from the fluxes of [O~{\sc iii}] lines, 
which were obtained with a high signal-to-noise ratio. The relatively high value of 
$T_{\rm e}$(O~{\sc iii}) is a consequence of the very low metallicity of J0811$+$4730.
We note that a temperature of $\ga$20000~K has been determined 
in some extremely metal-poor galaxies \citep[e.g. ][]{I09}. {\sc cloudy} 
models also predict the temperature range 20000 -- 23000 K
for objects with 12+logO/H $\sim$ 7.0, depending on the input parameters.

We derive an oxygen abundance of
12+logO/H = 6.98$\pm$0.02, making J0811$+$4730 the lowest-metallicity
SFG known, and the first galaxy with an oxygen abundance below 7.0.
We have also derived the oxygen abundance from the MODS1 and MODS2 spectra 
separately. We obtain respectively 12+logO/H = 6.98$\pm$0.03 and  6.97$\pm$0.03,
fully consistent with the value obtained from combining all data.
If we had adopted the higher extinction coefficient $C$(H$\beta$) = 0.185,  
derived from the hydrogen Balmer decrement by excluding the H$\alpha$ emission line, 
we would have obtained an electron temperature higher by 150 K  and a  slightly lower
oxygen abundance, 12+logO/H = 6.975$\pm$0.019. 
The N/O, Ne/O, S/O, Ar/O and Fe/O abundance ratios 
for this galaxy, shown in Table \ref{tab3}, are similar to those 
derived for low-metallicity SFGs \citep[e.g., ][]{I06}.
We note that the error of the electron temperature 
$T_{\rm e}$(O~{\sc ii}) has little impact on the error of 12+logO/H.
Adopting an error equal to 1500 K instead of 500 K in Table~\ref{tab3} does 
increase the error of O$^+$/H$^+$ by a factor of 3 but does not change much the 
error of 12+logO/H, increasing it by only 0.005 dex. This is because the O$^+$
abundance is several times lower than the O$^{2+}$ abundance. The largest impact
of such an increased error in  $T_{\rm e}$(O~{\sc ii}) is an increase in the errors of log N/O and 
log Fe/O by $\sim$ 0.03 dex. Similarly, increasing the $T_{\rm e}$(S~{\sc iii}) error from
500 K to 1500 K would result in increasing the errors in log S/O and log Ar/O 
by $\sim$ 0.03 dex.

\input{tab4.tex}

\section{Integrated characteristics of J0811$+$4730}
\label{sec:integr}

To derive the integrated characteristics of J0811$+$4730, we adopt 
the luminosity distance $D$ = 205 Mpc. It was obtained from the galaxy
redshift with the NED cosmological calculator 
\citep{W06}, adopting the cosmological parameters 
$H_0$ = 67.1 km s$^{-1}$ Mpc$^{-1}$, $\Omega_m$ = 0.318, $\Omega_\Lambda$ = 
0.682 \citep{P14} and assuming a flat geometry.
The absolute SDSS $g$ magnitude, corrected for the Milky Way
extinction $M_g$ = $-$15.41 mag (Table \ref{tab1}), characterizes J0811$+$4730 
as a dwarf SFG.

\subsection{H$\beta$ luminosity and star-formation rate}

To derive the total H$\beta$ luminosity $L$(H$\beta$), we need to correct the
observed H$\beta$ flux in the SDSS spectrum for extinction and spectroscopic 
aperture. Since the strong emission lines in the SDSS spectrum of J0811$+$4730
are clipped we adopt the extinction coefficient $C$(H$\beta$) = 0.165 obtained 
from the observed hydrogen Balmer decrement in the LBT spectrum. The aperture 
correction is determined as 2.512$^{g_{\rm ap}-g}$, where, $g$ = 21.38$\pm$0.05 mag 
and $g_{\rm ap}$ = 22.28$\pm$0.08 mag are respectively the SDSS total
modelled magnitude and the magnitude inside the spectroscopic aperture 
of 2 arcsec in diameter.
Both quantities, $g$ and $g_{\rm ap}$, are extracted 
from the SDSS data base. Using these data we obtain a correction factor of
2.3, which is applied to the H$\beta$ luminosity and some other 
quantities, such as the stellar mass $M_\star$, obtained from 
both SDSS and LBT spectra.
Here we have assumed that the spatial brightness distributions of 
the continuum and the hydrogen emission lines are the same. We have checked
this assumption by examining the LBT long-slit spectrum. 
We find from this spectrum that the brightness distributions of H$\beta$, 
H$\alpha$ and the adjacent continua are indeed nearly identical, with FWHMs of 
$\sim$ 0.6 arcsec, indicating that the galaxy 
is mostly unresolved. This angular size corresponds to an upper limit
of the galaxy linear size at FWHM of $\sim$ 600 pc.
 
The star-formation rate (SFR) in J0811$+$4730, 
derived from the aperture- and extinction-corrected H$\beta$ luminosity 
$L$(H$\beta$) using the \citet{K98} calibration, is 0.48~M$_\odot$~yr$^{-1}$.

\subsection{Stellar mass}




Fitting the spectral energy distribution (SED) of a galaxy is one of the most commonly 
used methods to determine its stellar mass \citep[e.g. ][]{CF05,As07}. The 
advantage of this method is that it can be applied to  
galaxies at any redshift and can separate stellar from nebular emission.
Here we derive the stellar mass of J0811$+$4730 from fitting the 
SEDs of both its SDSS and LBT spectra. 
Given the data at hand, it is the only method we can use.
The equivalent width of the
H$\beta$ emission line in both spectra of J0811$+$4730 is high,
EW(H$\beta$) $\sim$ 280\AA, indicating that star formation occurs in a burst
and that the contribution of the nebular continuum is also high and should be 
taken into account in the SED fitting, in addition to the stellar emission. 
In particular, the fraction of the nebular continuum near H$\beta$ is 
$\sim$ 30 per cent, and it is considerably higher near H$\alpha$.
The details of the SED fitting are described e.g. in \citet*{I11} and 
\citet{I15}. The salient features are as follows.

We carried out a series of Monte Carlo simulations to reproduce  the SED of
J0811$+$4730. To calculate the contribution of the  stellar  emission to the 
SEDs, we adopted the grid of the Padua stellar evolution models by \citet{G00}
with heavy element mass fraction $Z$ = 0.0004, or 1/50 solar. To reproduce the 
SED of the stellar component with any star-formation history, we used the 
{\sc Starburst99} models \citep{L99,L14} 
to calculate a grid of instantaneous
burst SEDs for a stellar mass of 1 M$_\odot$ in a wide range of ages
from 0.5 Myr to 15 Gyr, and \citet*{L97} stellar atmosphere models. 
We adopted a stellar initial mass function with a 
Salpeter slope \citep{S55}, an upper mass limit of 100 M$_\odot$, and a lower 
mass limit of 0.1 M$_\odot$. Then the SED with any star-formation history can 
be obtained by integrating the instantaneous burst SEDs over time with a 
specified time-varying SFR.

We approximated the star-formation history in J0811$+$4730
by a recent short burst forming young stars at age $t_y$ $<$ 10 Myr and a 
prior continuous star formation 
with a constant SFR responsible for the older stars with ages ranging from 
$t_2$ to $t_1$, where $t_2 > t_1$ and varying between 10 Myr and 
15 Gyr. Zero age is now. The contribution of each stellar population to the 
SED was parameterized by the ratio of the masses of the old to young stellar 
populations, $b=M_y/M_o$, which we varied between 0.01 and 100.

The total modelled  monochromatic (nebular and stellar)
continuum flux near the H$\beta$ emission line for a mass of 1 M$_\odot$
was scaled to fit the monochromatic extinction- and aperture-corrected 
luminosity of the galaxy at the same wavelength. The
scaling factor is equal to the total stellar mass $M_\star$ in solar units.
In our fitting model, $M_\star=M_y+M_o$, where $M_y$ and $M_o$ were derived 
using $M_\star$ and $b$.

\begin{figure}
\includegraphics[angle=-90,width=0.99\linewidth]{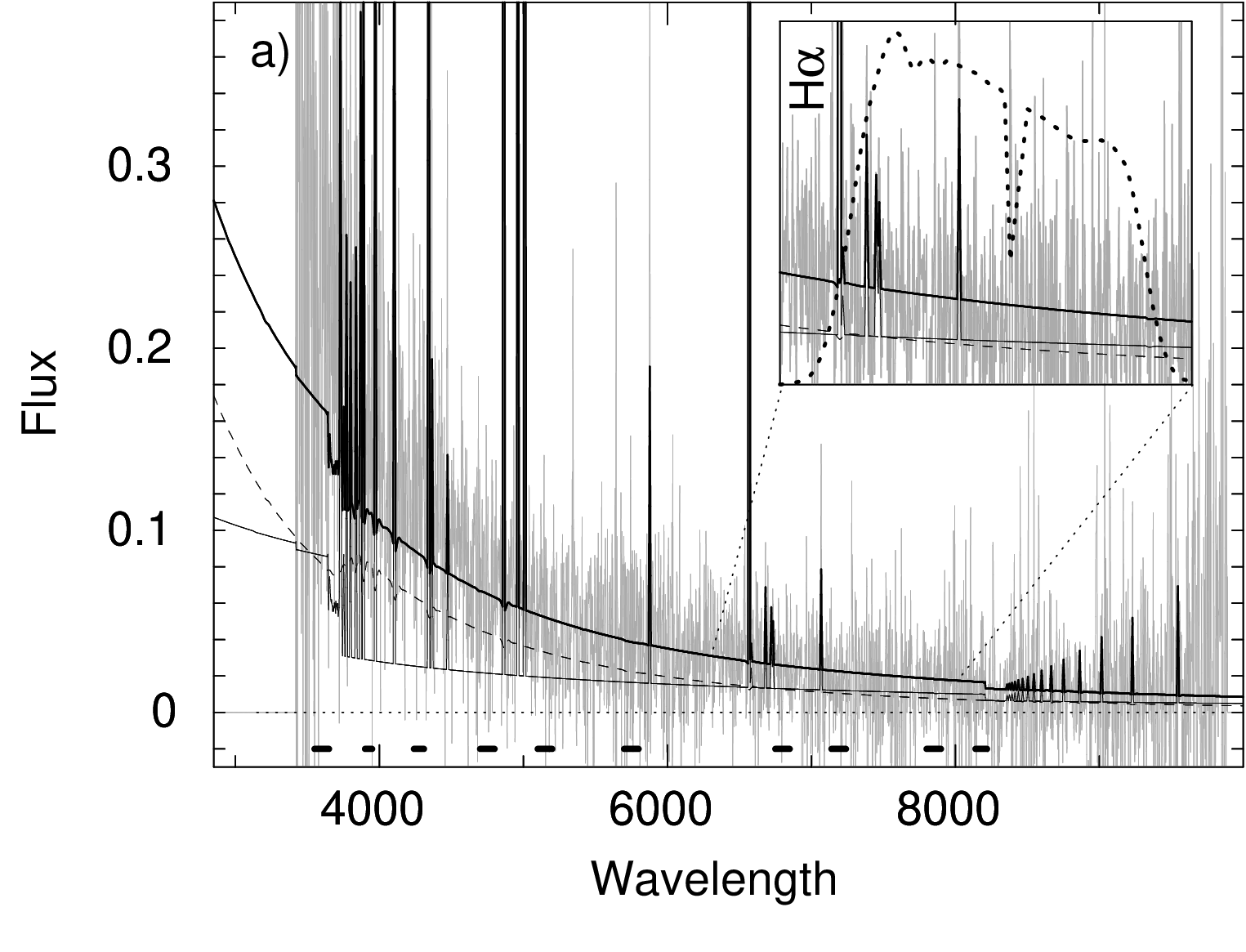}
\includegraphics[angle=-90,width=0.99\linewidth]{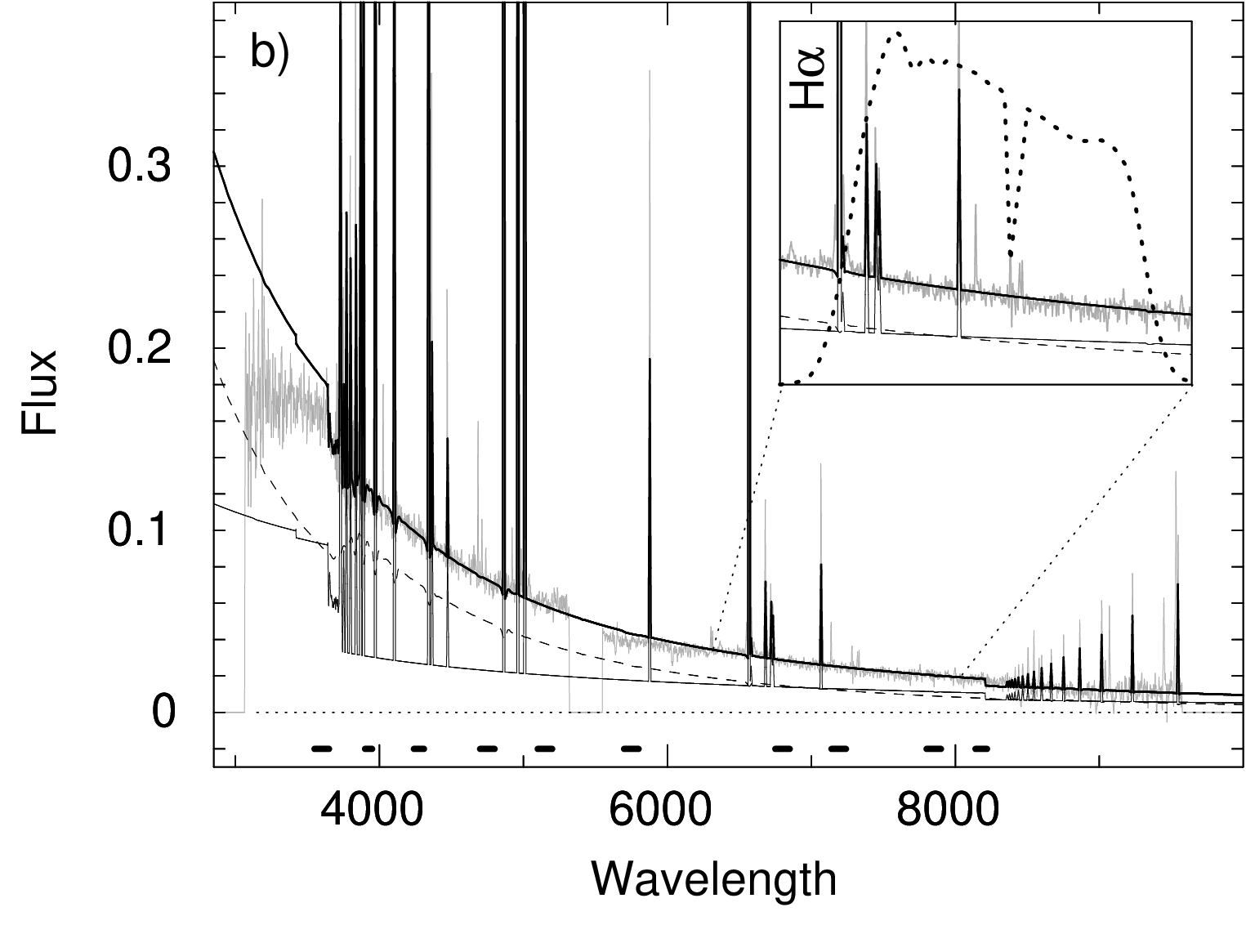}
\caption{ {\bf a)} The rest-frame extinction-corrected SDSS spectrum of 
J0811$+$4730
overlaid by the modelled nebular SED (thin solid line), stellar SED (dashed
line), and total (stellar and nebular) SED (thick solid line).
Short horizontal lines indicate wavelength ranges used for SED fitting.
The inset shows the expanded part of the spectrum that falls into the SDSS $i$ 
photometric band. The dotted line in the inset is the transmission curve of the
SDSS $i$ filter. Wavelengths are in \AA\ and fluxes are in units of  
10$^{-16}$ erg s$^{-1}$ cm$^{-2}$ \AA$^{-1}$.
{\bf b)} Same as in {\bf a)}, but the rest-frame extinction-corrected LBT 
spectrum of J0811$+$4730 and SED fits to it are shown. The meaning of the lines is the
same as in {\bf a)}.}
\label{fig2}
\end{figure}

The SED of the nebular continuum was taken from \citet{A84}. It
included hydrogen and helium free-bound, free-free, and two-photon emission. 
In our models, this was always calculated with
the electron temperature $T_{\rm e}$(H$^+$) of the H$^+$ zone, which is not 
necessarily equal to $T_{\rm e}$(O~{\sc iii}). We thus vary it in the range
(0.95 -- 1.05)$\times$$T_{\rm e}$(O~{\sc iii}). The fraction of the nebular
continuum near H$\beta$ is defined as the ratio of the observed EW(H$\beta$)
to the pure nebular value, which is $\sim$ 900 -- 1000\AA, depending of the
electron temperature. The observed intensities
of the emission lines from the LBT spectrum were corrected for reddening and 
scaled using the flux of the H$\beta$ 
emission line and were added to the calculated nebular continuum. 
However, these lines were not used in the SED fitting, because we fit only the 
continuum. 
We also varied the extinction coefficient 
$C$(H$\beta$)$_{\rm SED}$ in the range of 
(0.9 -- 1.1)$\times$$C$(H$\beta$), where $C$(H$\beta$) = 0.165, the extinction 
coefficient derived from the observed hydrogen Balmer decrement in the LBT
spectrum.
  
For J0811$+$4730, we calculated 5$\times$10$^5$ Monte Carlo models 
for each of the SDSS and LBT spectra by 
varying $t_y, t_1, t_2, b$, and $C$(H$\beta$)$_{\rm SED}$. The best model should 
satisfy three conditions. First, the modelled EW(H$\beta$) should match 
the observed value, within the adopted range between 0.95
and 1.05 times its nominal value. Second, the modelled EW(H$\alpha$) 
should also match the observed value, in the same range between 0.95
and 1.05 times its nominal value. As compared to our previous work, this condition is introduced here 
for the first time as it would better constrain the models.
Third, among the models satisfying 
the first and the second conditions, the best 
modelled SEDs of the SDSS and LBT spectra were found 
from $\chi^2$ minimization of the deviation between the 
modelled and the observed continuum at ten wavelength ranges, which are 
selected to cover the entire spectrum and to be free 
of emission lines and residuals of night sky lines.

From the fit of the SDSS spectrum, we find that the best model for the stellar 
population consists of young
stars with an age $t_y$ = 3.28 Myr and a stellar mass 
$M_y$ = 10$^{6.13}$ M$_\odot$, and of older stars formed continuously with 
a constant SFR over the time period from 770 Myr to 930 Myr and a stellar 
mass $M_o$ = 10$^{5.59}$ M$_\odot$, or 3.5 times lower than the stellar mass of 
the young stellar population. Correspondingly, the total stellar mass of the 
galaxy is $M_\star$ = $M_y$ + $M_o$ = 10$^{6.24\pm0.33}$~M$_\odot$. 
This implies that 78\% of the total stellar mass has been formed during the 
most recent burst of star formation.
Similarly, from the fit of the LBT spectrum, we derive 
$M_y$ = 10$^{6.18}$ M$_\odot$ for the young population with an age 
$t_y$ = 3.28 Myr, and a mass $M_o$ = 10$^{5.63}$ M$_\odot$ of older stars formed
continuously with a constant SFR over the time period from 100 Myr to 290 Myr.
This corresponds to a total stellar mass $M_\star$ = 10$^{6.29\pm0.06}$~M$_\odot$, in
agreement, within the errors, with the value obtained from fitting the SDSS 
spectrum. 
All masses obtained from the SDSS spectrum have been corrected for aperture effects 
using the SDSS $g$-band total magnitude and the 
$g$ magnitude inside the 2 arcsec SDSS aperture. This correction would be somewhat different 
for the LBT spectrum obtained with a 1.2 arcsec wide slit. We do not have enough data to determine the 
LBT aperture correction accurately, so for the sake of simplicity, we have 
adopted the same aperture correction for the LBT data as for the SDSS data.
The good agreement between the SDSS and LBT stellar masses suggests that this 
procedure is not too far off. 

We note that the determination
of the age and the mass of the old stellar population is somewhat uncertain
because its contribution to the continuum in the optical range is low, not
exceeding 6 per cent near H$\beta$ and 7 per cent near H$\alpha$ in the best
model. However, even if we fix the formation period of the old stellar 
population to be between 1 Gyr and 10 Gyr ago, i.e. we assume the least luminous
old stellar population with the highest mass-to-luminosity ratio, we still find 
from SED fitting a small total stellar mass $M_\star$ $\sim$ 10$^{6.34}$ M$_\odot$.
Moreover, setting additionally $C$(H$\beta$) = 0 in the SED fitting does not 
change the conclusion about the low galaxy stellar mass.

The rest-frame extinction-corrected SDSS and LBT spectra overlaid by the 
SEDs of the best models are shown in Fig.~\ref{fig2}a and \ref{fig2}b,
respectively. The nebular, stellar and total (nebular plus stellar) SEDs are 
shown respectively by thin solid, dashed and thick solid lines. 
We use only regions clean of galaxy
emission lines and night-sky residuals to fit the SEDs. They are shown in 
Fig.~\ref{fig2} by short horizontal lines.

We find 
that the contribution of the nebular continuum near the H$\beta$ emission line 
is nearly 32 per cent of the total continuum emission in that region. 
Furthermore, the SED of the nebular continuum is shallower than 
the stellar continuum SED (compare the thin solid and dashed lines in 
Fig.~\ref{fig2}).
This makes the fraction of nebular continuum
increase to $\sim$50 per cent of the total near the H$\alpha$ emission line.
Therefore, neglecting the nebular continuum, 
i.e. assuming that the continuum shown by the thick solid line in 
Fig.~\ref{fig2} is purely stellar in origin, would result in an older
stellar population and a larger stellar mass. Indeed, with this assumption, we
find a stellar mass which is 0.56 dex higher than the one derived above.
This overestimate of the stellar mass is consistent with the finding of \citet{I11} 
who compared the stellar mass determinations for similar compact SFGs, dubbed "Green Peas",
with the method described in this paper and the method used by \citet{Ca09},
which does not take into account nebular continuum emission. \citet{I11} found that 
the neglect of the nebular continuum contribution would result in an overestimate of the stellar masses of 
compact SFGs with EW(H$\beta$) $\geq$ 100\AA\ by 0.4 dex on average.

In principle, the stellar mass of the old stellar population in galaxies 
can be estimated from the SDSS $i$ magnitude, if all light in this
band is stellar and assuming a value for the  
mass-to-light ratio which for the oldest stars is $\sim$ 1, if the mass and 
luminosity are expressed in solar units. However, this technique does not  
work for J0811$+$4730 because the contribution of the cool and
old stars to its $i$-band luminosity is small. To demonstrate this, we
show in insets of Fig.~\ref{fig2} the parts of the spectra that fall into the SDSS 
$i$-band, with the dotted lines representing the $i$-band transmission curve, with a full passband width FWHM $\sim$1200\AA. 
It is seen that the 
H$\alpha$ emission line, with a EW(H$\alpha$) of $\sim$ 1700\AA, is redshifted to a
wavelength where the sensitivity of the transmission curve is still $\sim$1/3 of 
its maximum value. Therefore, the H$\alpha$ line contributes, within the  
spectroscopic aperture, 1/3$\times$EW(H$\alpha$)/(1/3$\times$EW(H$\alpha$) + 
FWHM) $\sim$ 1/3 of the light in the  $i$-band. 
Furthermore, to account for the high value of EW(H$\alpha$), about half of the
remaining light must be due to the nebular continuum (compare the dashed and 
thin solid lines in the insets of Fig.~\ref{fig2}). This nebular continuum 
comes from hydrogen recombination 
and free-free emission and cannot be 
neglected. We estimate that only $\sim$ 1/3 of the J0811$+$4730 light in the 
$i$-band is of stellar origin. Furthermore, a major fraction of this stellar emission comes from 
the Rayleigh-Jeans tail of the radiation distribution of hot and luminous young stars.
We thus conclude that only a small fraction of the $i$-band light comes from 
the old stellar population and that photometry 
cannot be used for simple mass estimates of the old stellar population in 
J0811$+$4730. 
Only SED fitting of spectra, which includes both the stellar and 
nebular emission in a wide range of wavelengths, can give physically 
justified estimates.

The low stellar mass $M_\star$, in addition to its faint absolute
magnitude, characterizes J0811$+$4730 as a dwarf SFG.

\begin{figure}
\centering{
\includegraphics[angle=-90,width=0.98\linewidth]{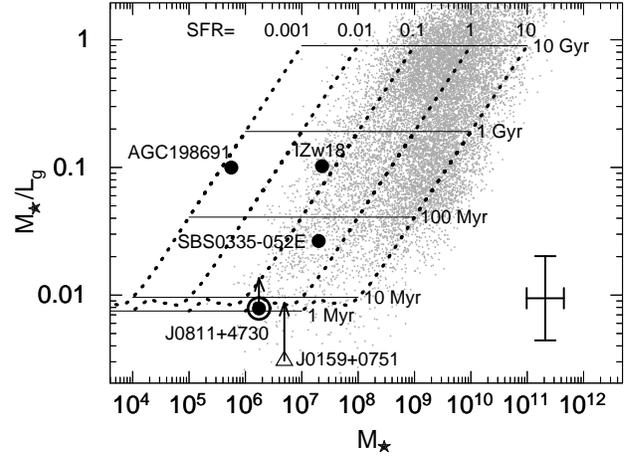}}
\caption{Relation between the stellar mass-to-luminosity ratio
and the stellar mass. All quantities are corrected for spectroscopic
aperture and extinction, and are expressed in solar units.
Selected lowest-metallicity SFGs with 12+logO/H$<$7.3 are represented
by labelled filled circles. The galaxy J0811$+$4730 is encircled. 
The error bar indicates the errors of $M_\star$ and $M_\star$/$L_g$ for this SFG.
By an open triangle is shown the SFG J0159$+$0751, characterized by very high
O$_{32}$ = [O~{\sc iii}]$\lambda$5007/[O~{\sc ii}]$\lambda$3727 = 39 and
EW(H$\beta$) = 347\AA\ \citep*{I17}. Arrows indicate the upward shifts of the 
latter SFG and J0811$+$4730 if the contribution of the H$\beta$ and 
[O~{\sc iii}]$\lambda$4959, $\lambda$5007 emission in the SDSS $g$-band 
is excluded. For comparison, 
are also shown compact SFGs from the SDSS DR12 with redshifts $z$$>$0.01
\citep[grey dots, ][]{I16c}. Dotted lines represent {\sc Starburst99} models 
with continuous star formation and a SFR varying from 
0.001 to 10  M$_\odot$ yr$^{-1}$, during the periods from the present to 10 Gyr, 
1 Gyr, 100 Myr, 10 Myr, and 1 Myr in 
the past (horizontal solid lines). These models include the contribution of
both the stellar and nebular continua, but not of nebular emission lines.}
\label{fig3}
\end{figure}

\subsection{Comparison with I Zw 18}

We test the robustness of our method of stellar mass determination by applying it to 
another extremely metal-poor SFG, I Zw 18. This SFG has been selected because
its stellar mass can be determined by at least another independent method.
\citet{A13} have used the CMD of its resolved stellar populations 
to derive a stellar mass of $>$10$^{7.24}$ M$_\odot$ of its main body, adopting
a distance of 18.2 Mpc \citep{A07}. 

This is to be compared with the stellar mass obtained from the SED fitting
described above. We use the SDSS spectra of the two brightest NW and SE 
components constituting the main body of I Zw 18. We obtain aperture-corrected
stellar masses 10$^{7.34}$ M$_\odot$ and 10$^{6.30}$ M$_\odot$ for the NW and SE
components, respectively. The total stellar mass of the I Zw 18 main body derived by SED fitting is therefore 10$^{7.38}$ M$_\odot$, consistent with the lower limit derived by \citet{A13} from the CMD.

\subsection{Mass-to-luminosity ratio}

In Fig. \ref{fig3} we show the stellar mass-to-luminosity 
ratio vs. stellar mass diagram for some of the most metal-poor SFGs known (filled circles). The galaxy
J0811+4730 is encircled. For three SFGs (J0811$+$4730, AGC~198691 and I~Zw~18),
we use SDSS $g$-band luminosities
calculated from the modelled total apparent magnitudes and adopting the 
absolute $g$-band magnitude of 5.45 mag for the Sun \citep{B03}.  
We note that the total magnitude used for I Zw 18 includes both the SE 
and NW components. The photometric
data for SBS~0335$-$052E are not present in the SDSS data base.
We have therefore adopted for it the $B$ magnitude given by \citet*{P04}, taking the absolute
$B$ magnitude of the Sun to be 5.48 \citep{BM98}. 
We have used the stellar masses of AGC~198691 and SBS~0335$-$052E derived by
\citet{H16} and \citet{I14}, respectively. The stellar masses of I~Zw~18 
and of J0811$+$4730 are derived in this paper by SED fitting
of their SDSS spectra. 

The mass-to-luminosity ratio of J0811$+$4730 is extremely
low, $\sim$ 1/100, as derived from both the SDSS and LBT SEDs,
or approximately one order of magnitude lower than the 
mass-to-luminosity 
ratio of the second most metal-poor galaxy known, SFG AGC~198691 \citep{H16}, 
and of I~Zw~18 (Fig. \ref{fig3}), but only a factor of 3 lower than that of 
another extremely metal-deficient SFG, SBS~0335$-$052E. 
For comparison, we show also in Fig. \ref{fig3} by grey dots the compact SFGs in
the SDSS DR12 \citep{I16c}. The stellar masses for these galaxies are derived 
from the SEDs, using the same technique as described above. It is evident that, 
compared to  J0811$+$4730, the majority of the compact SFGs in  
the SDSS DR12 are characterised by much higher
$M_\star$/$L_g$ ratios. However, there is a 
small, but non-negligible number of SFGs with mass-to-luminosity 
ratios similar to that of J0811$+$4730. They are characterized by high 
equivalent widths EW(H$\beta$) ($>$ 200\AA),  as derived from their SDSS 
spectra.

To study and compare the evolutionary status of compact SFGs, we
present in Fig. \ref{fig3} {\sc Starburst99} continuous star
formation models with a constant SFR, occurring from a specified time $t$ 
in the past to the present. For the sake of definiteness, models have been 
calculated for a metallicity of 1/20 solar. They are shown by dotted lines, 
labelled by their SFRs which vary from 0.001 to 10 M$_\odot$ yr$^{-1}$. 
We indicate by horizontal
solid lines the $M_\star$/$L_g$ ratios corresponding to $t$ = 1 Myr, 10 Myr, 
100 Myr, 1 Gyr and 10 Gyr. The behavior of models for other metallicities is 
similar to that represented in Fig. \ref{fig3}.
It is seen in the Figure that for $t$ $\leq$ 10 Myr,
corresponding to a burst model, the $M_\star$/$L_g$ ratio is 
very low, $\sim$ 0.01, and nearly constant, with a value 
consistent with the one for an instantaneous burst \citep{L99}. For larger
$t$, the $M_\star$/$L_g$ ratio increases with stellar mass, 
following the relation $\sim$ $M_\star^{2/3}$, with a corresponding increase 
of the mass fraction of old stars, approximately following the relation
[$M_\star$($t$) -- $M_\star$($\leq$10 Myr)]/$M_\star$($t$). It is interesting to
note that the distribution of compact SFGs from the SDSS DR12 (grey dots in 
Fig. \ref{fig3}) follows these relations, with an increasing mass fraction of 
stars with age $>$ 10 Myr in more massive compact SFGs, in agreement with the
conclusion of \citet{I11}.

According to our SED fitting, a considerable fraction of the stellar mass 
in J0811$+$4730 was created during the last burst of star formation.
As for the other very metal-poor SFGs shown in Fig. \ref{fig3},
the mass fraction of older stars is higher. According to the locations
of I~Zw~18, SBS~0335$-$052E and AGC~198691 in this diagram, older stars with 
age $>$ 10 Myr dominate the stellar mass in those galaxies. 

The models in Fig. \ref{fig3} include both the stellar and nebular
continua. However, we note the important contribution of the 
H$\beta$ and [O~{\sc iii}]$\lambda$4959, $\lambda$5007 emission lines to  
$L_g$ in SFGs with a very high EW(H$\beta$). Therefore, to compare the $M_\star$/$L_g$ of those SFGs with model predictions, the contribution of the emission
lines should be subtracted from the $L_g$'s derived from 
their $g$-band magnitudes. The magnitude of the effect depends on the EW(H$\beta$) and the
metallicity. To demonstrate this, we show in Fig. \ref{fig3} by an open
triangle the location of the compact SFG J0159$+$0751 with a $M_\star$/$L_g$
ratio far below the model predictions. This SFG, with an 
oxygen abundance 12+logO/H = 7.55, is characterised by
a very high [O~{\sc iii}]$\lambda$5007/[O~{\sc ii}]$\lambda$3727 flux
ratio of 39 and an EW(H$\beta$) = 347\AA, indicative of  a very young starburst 
\citep*{I17}. However, subtracting the nebular line emission would put J0159$+$0751
in the region allowed by models (shown by an upward arrow in Fig. \ref{fig3}). 
Excluding nebular line emission results in a more modest upward
shift for J0811$+$4730, also shown by an arrow. The shift is smaller because of
a much lower metallicity and thus a much fainter 
[O~{\sc iii}]$\lambda$4959, $\lambda$5007 emission. For the other most 
metal-poor
SFGs shown in Fig. \ref{fig3}, the contribution of nebular emission lines is
even lower because of their considerably lower EW(H$\beta$).

\section{Emission-line diagnostic and metallicity-luminosity diagrams}
\label{sec:diagrams}

\begin{figure}
\centering{
\includegraphics[angle=-90,width=0.98\linewidth]{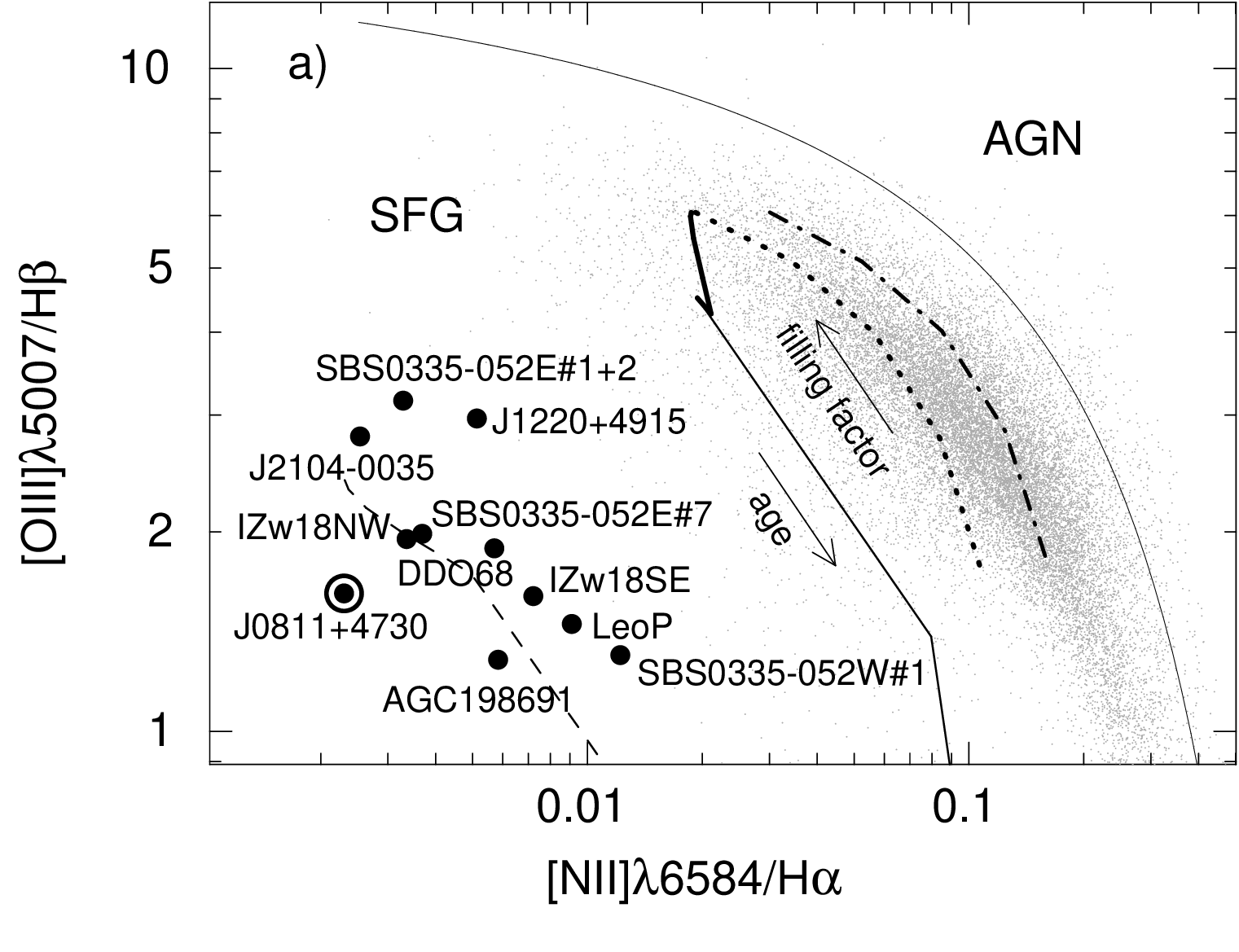}
\includegraphics[angle=-90,width=0.98\linewidth]{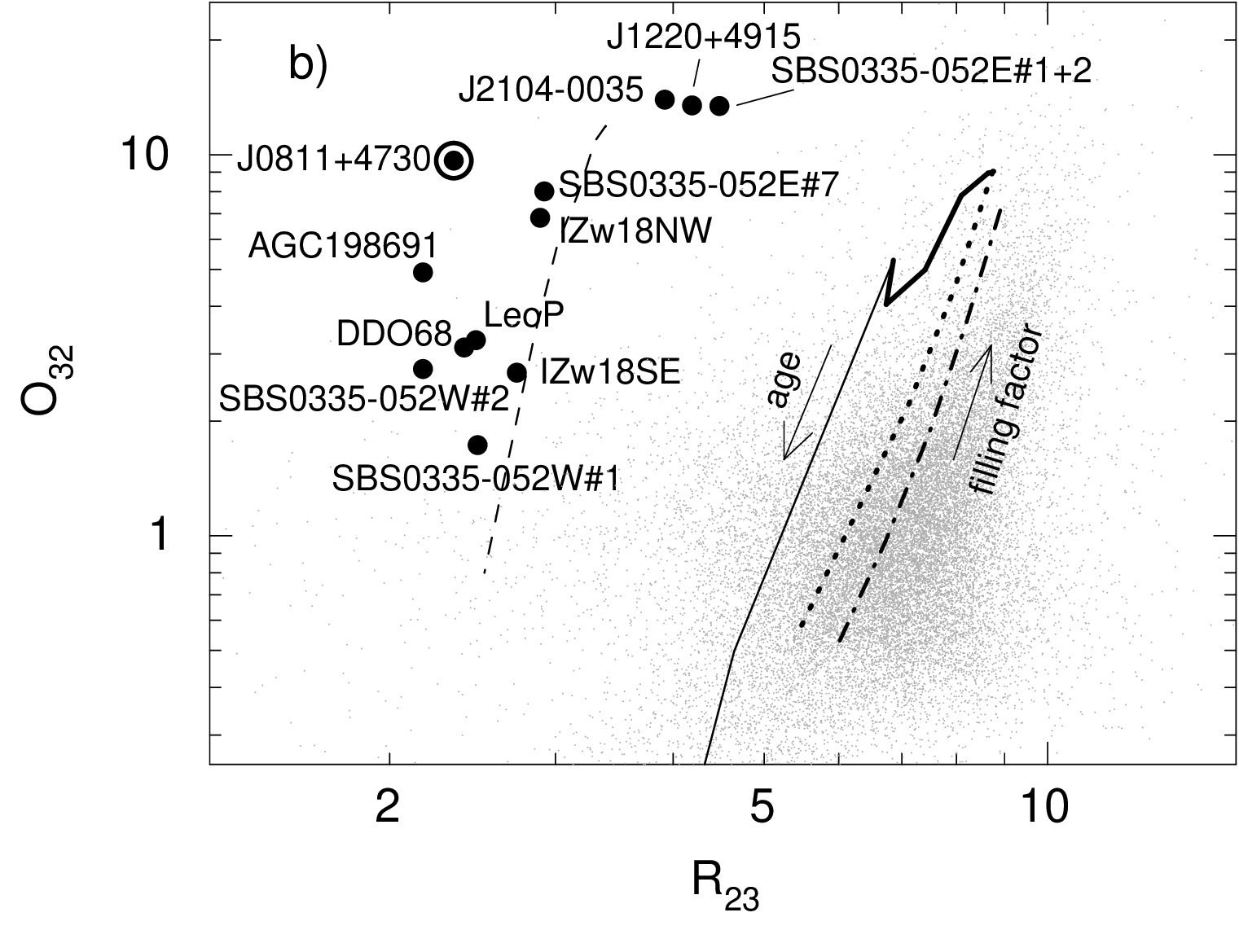}
}
\caption{ {\bf (a)} The [O~{\sc iii}]$\lambda$5007/H$\beta$ -- 
[N~{\sc ii}]$\lambda$6584/H$\alpha$ diagnostic diagram 
\citep*[BPT;][]{BPT81}. The lowest-metallicity SFGs with 12+logO/H$<$7.3
are shown by filled circles. The galaxy J0811$+$4730 is encircled.
For comparison, are shown compact SFGs from the SDSS DR12
\citep[grey dots, ][]{I16c}. The thin solid line by \citet{K03} 
separates SFGs from active galactic nuclei (AGN). 
Dashed and thick solid lines represent relations obtained for
{\sc cloudy} photoionized H~{\sc ii} region models with 12+logO/H = 7.3 and
8.0, respectively, and with different starburst 
ages ranging from 0 Myr to 6 Myr. Models with 12+logO/H = 8.0 and 
starburst ages of 0 - 4 Myr, corresponding to the highest H$\beta$ luminosity 
and equivalent width are represent by a thicker solid line.
The model dependences with a zero
starburst age but with a varying filling factor in the range 10$^{-1}$ -- 
10$^{-3}$, for two oxygen abundances 12+logO/H = 8.0 and 8.3, are shown by dotted
line and dash-dotted line, respectively. The directions of increasing age 
and filling factor are indicated respectively by downward and upward arrows.
{\bf (b)} The O$_{32}$ - R$_{23}$ diagram for SFGs, where 
O$_{32}$=[O~{\sc iii}]$\lambda$5007/[O~{\sc ii}]$\lambda$3727 and 
R$_{23}$=([O~{\sc ii}]$\lambda$3727 +
[O~{\sc iii}]$\lambda$4959 + [O~{\sc iii}]$\lambda$5007)/H$\beta$.
The meaning of lines and symbols is the same as in (a). 
}
\label{fig4}
\end{figure}

\citet{I12} and \citet{G17} have demonstrated that the lowest-metallicity
SFGs occupy a region in the diagram by \citet*{BPT81} (hereafter BPT), that is 
quite different from the location of the 
main-sequence defined by other nearby SFGs. 
This fact is clearly shown in Fig.~\ref{fig4}a.
The data for the lowest-metallicity SFGs in this Figure are 
presented in Table \ref{tab4}. The references in the Table are for the
oxygen abundances.
The most metal-deficient SFG known, 
J0811$+$4730, is shown as an encircled filled circle in this diagram.
It is the most outlying
object among the sequence defined by the lowest-metallicity SFGs 
(filled circles) which is itself much shifted to the left from the ``normal'' 
SFG main-sequence (grey dots).

\begin{figure}
\centering{
\includegraphics[angle=-90,width=0.98\linewidth]{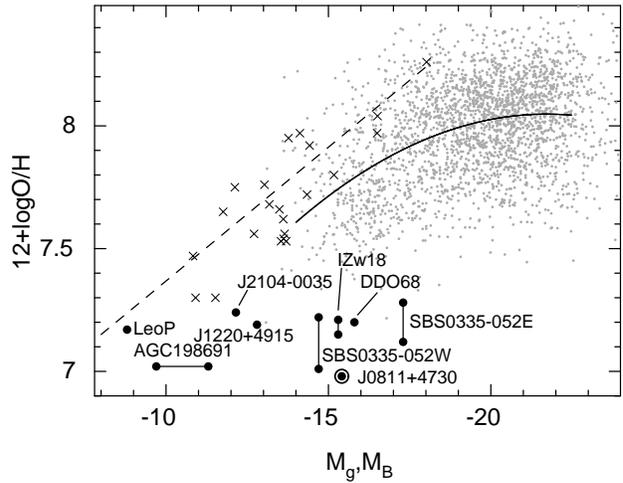}
}
\caption{The oxygen abundance -- absolute magnitude diagram. Low-luminosity
galaxies (crosses), with the fit to these data by \citet{B12} (dashed line).
The solid line is the fit to the SDSS compact SFGs from DR12 with oxygen 
abundances derived by the direct $T_{\rm e}$ method \citep[grey dots, ][]{I16c}.
Other symbols are the same as in Fig. \ref{fig4}.
Vertical solid lines indicate the ranges of oxygen abundance in different 
star-forming regions in I Zw 18, SBS 0335$-$052W and SBS 0335$-$052E. 
Horizontal line connects the positions of AGC~198691, calculated 
with two distances, 8 and 16 Mpc.}
\label{fig5}
\end{figure}

The shift of the sequence  
to the left of the BPT diagram is caused by lower metallicities.
To demonstrate this, we plot the [O~{\sc iii}]$\lambda$5007/H$\beta$ -- 
[N~{\sc ii}]$\lambda$6584/H$\alpha$ relations as derived from photoionization
H~{\sc ii} region models for two oxygen abundances, 12+logO/H = 7.3 (dashed 
line) and 8.0 (solid line), and for various ages of the 
starburst, ranging from 0 to 6 Myr. The direction of age increase is
indicated by the downward arrow. The models are calculated with the 
{\sc cloudy} code c13.04 \citep{F98,F13}, adopting a production rate
of ionizing radiation $Q$ = 10$^{53}$ s$^{-1}$ and a filling factor 
$f$ = 10$^{-1}$. Since the location of the theoretical relations depends on the
adopted input N/O abundance ratio, we have adopted logN/O = $-$1.5 for models 
with 12+logO/H = 7.3, and logN/O = $-$1.2 for models with 12+logO/H = 8.0, 
values that are typical of low-metallicity SFGs \citep[e.g. ][]{I06}. 
The part of the modelled dependence with 12+logO/H = 8.0 and starburst
ages 0 -- 4 Myr, corresponding to the stages with highest H$\beta$ luminosities
and equivalent widths, is shown by a thicker solid line. \citet{I16c} selected
the SDSS DR12 compact SFGs (grey dots) to have high EW(H$\beta$) and thus they 
correspond to this part of the modelled sequence.
 
We note however that the location of the modelled sequences in 
Fig. \ref{fig4} depends on the adopted input parameters and therefore they
are shown only to illustrate dependences on metallicities. The location of 
main-sequence SFGs is closest to models with oxygen abundance 
12+logO/H = 8.0. 
For comparison, we also show the modelled dependences with a fixed 
starburst age of 0 Myr, but varying the filling factor 
in the range 10$^{-1}$ -- 10$^{-3}$ for oxygen abundances 12~+~logO/H = 8.0 
(dotted line) and 8.3 (dash-dotted line). The effect of an increasing filling factor is shown by an upward arrow. These models reproduce much better
the location of the main-sequence SFGs (grey dots). This is to be expected 
since \citet{I16c}
selected from the SDSS DR12 only compact SFGs with high EW(H$\beta$), as noted
above, i.e. with young bursts.
On the other hand, the location of the lowest-metallicity 
SFGs agrees well with models calculated for an oxygen abundance 
12+logO/H = 7.3. 

Similarly, the lowest-metallicity SFGs are located in the O$_{32}$ -- R$_{23}$ 
diagram in a region which is very different from that of the main-sequence of 
the SDSS SFGs (Fig. \ref{fig4}b), with J0811$+$4730 being one of the 
galaxies  most shifted to the left (encircled filled circle). 
These shifts to the left are again primarily due to lower metallicities.
This is seen by comparing the 
predictions of models with varying starburst age for 
12+logO/H = 7.3 (dashed line) and
8.0 (solid line). It is also seen that models for zero-age starbursts 
with varying filling factors for 12+logO/H = 8.0 (dotted line) and
8.3 (dash-dotted line) reproduce much better the location of the main-sequence
galaxies (grey dots) as compared to the modelled sequence of bursts with a 
varying age (solid line).

We emphasize that the modelled emission-line
ratios in Fig. \ref{fig4} have been calculated taking into account only stellar
ionizing radiation. These ratios may be changed if additional sources of 
ionization and heating such as shocks and X-ray emission are present.
Inclusion of these additional sources would enhance 
the [O~{\sc ii}]$\lambda$3727 emission line relative  
to the [O~{\sc iii}]$\lambda$5007 emission line \citep[e.g. ][]{S15}, reducing 
O$_{32}$ and shifting the modelled sequences downward.

Finally, in Fig. \ref{fig5} we present the oxygen abundance -- absolute 
magnitude relation for SFGs. 
For some galaxies, absolute 
$B$-band magnitudes have been adopted, while for others absolute SDSS $g$-band
magnitudes have been used. This difference in the adopted magnitudes 
will change little the distributions of SFGs in 
Fig. \ref{fig5} because the differences in $g$ and $B$ magnitudes 
for SFGs are small, $\la$ 0.1 mag. We also note that $M_B$ for AGC~198691 
in Table \ref{tab4} is derived assuming a distance of 16 Mpc.

Oxygen abundances for
all galaxies shown in this Figure have been derived by the direct $T_{\rm e}$ 
method.
In particular, only compact SFGs from the SDSS DR12 (grey dots) 
with an [O~{\sc iii}]$\lambda$4363 emission line detected in their spectra
with an accuracy better than 50 per cent, allowing a reasonably good determination
of 12+logO/H by the direct method, are included. Following \citet{I15}, we stress the 
importance of using the same method of metallicity determination for all 
data, to exclude biases introduced by different methods and to produce an 
homogeneous set of data. In particular,
\citet{I15} showed that, concerning the often cited oxygen abundance -- absolute 
magnitude and oxygen abundance -- stellar mass relations derived for SFGs by 
\citet{T04}, the metallicities
derived by strong-line methods can be by as much as 0.5 dex 
higher compared to the metallicities derived by the direct method for the
same SFGs. They are inconsistent with those derived by other strong-line 
methods 
such as those proposed e.g. by \citet{PP04}, \citet{PT05} and \citet{I15}. 
Those authors  
calibrated their relations using oxygen abundances derived by the direct 
method, and the abundances obtained from their calibrations are thus 
consistent with the $T_{\rm e}$ method.

The lowest-metallicity SFGs present a wide range of absolute magnitudes. The 
two galaxies with the lowest luminosities are Leo P
\citep{S13} and AGC~198691 \citep{H16}. The first galaxy follows well  
the relation found by \citet{B12} for relatively quiescent 
SFGs, while the second one deviates somewhat from it to a higher luminosity.

On the other hand, some other lowest-metallicity SFGs, including 
J0811$+$4730, strongly deviate from the \citet{B12} relation.
\citet{I12} and \citet{G17} have attributed these deviations 
to the enhanced brightnesses of the galaxies undergoing  
active star formation. In particular, we estimate that the luminosity 
increase for J0811$+$4730
with its high EW(H$\beta$) would be as much as 2 magnitudes, due to the 
contribution of the nebular continuum and emission lines to the $g$-band 
luminosity. However, such an increase is insufficient to
explain the whole deviation.

Additionally, these galaxies can also be  
chemically unevolved objects with a short star formation history,  
having too low metallicities for their high luminosities. Indeed, 
star formation in a system such as SBS~0335$-$052E is very different from that
in dwarfs like Leo~P and AGC~198691. It is currently undergoing a powerful 
burst, giving birth to bright 
low-metallicity super-star clusters containing thousands of 
massive O-stars \citep*{TIL97}, followed by a delayed element enrichment. 
A similar scenario can be at work in J0811$+$4730 and at high redshifts, in 
zero-metallicity
dwarf primeval galaxies, undergoing their first burst of star formation.

Finally, \citet{EC10} have proposed to explain these low metallicities by 
the infall of gas from galactic halos. The effective mixing
of the metal-poor gas in the halos with the more metal-rich 
gas in the central part of the galaxies will dilute the latter, decreasing its metallicity. Clearly more 
observational and modelling work is needed to understand the true nature of J0811$+$4730.

\section{Conclusions}\label{sec:conclusions}

In this paper we present Large Binocular Telescope (LBT) 
spectrophotometric observations of the compact star-forming galaxy (SFG)
J0811$+$4730 selected from the Data Release 13 (DR13) of the Sloan Digital Sky
Survey (SDSS). We find that the oxygen abundance of this galaxy is 12+logO/H 
= 6.98$\pm$0.02, the lowest ever found for a nearby SFG, and the first one 
below 12+logO/H = 7.0. J0811$+$4730 strongly deviates from the SDSS
main-sequence SFGs in the [O~{\sc iii}]$\lambda$5007/H$\beta$ -- 
[N~{\sc ii}]$\lambda$6584/H$\alpha$ and 
[O~{\sc iii}]$\lambda$5007/[O~{\sc ii}]$\lambda$3727 -- 
([O~{\sc ii}]$\lambda$3727 + [O~{\sc iii}]$\lambda$4959 + 
[O~{\sc iii}]$\lambda$5007)/H$\beta$ diagrams because of its extremely low 
metallicity. In the same way as other galaxies with very low metallicities, it 
is also strongly offset in the oxygen abundance -- absolute magnitude 
diagram from the relations defined by nearby galaxies \citep{B12} and 
compact SFGs \citep{I16c}. This offset can probably be explained by a 
combination of its chemically unevolved nature, 
an enhanced brightness of its star-forming regions, 
and gas infall resulting in the dilution of the more metal-rich gas in 
the inner region by the outer more metal-poor gas in the halo.

The properties of the local lowest-metallicity SFGs are likely close to those 
of the high-redshift low-luminosity SFGs recently found at $z$ $>$ 3
\citep{K17}, and thought to have played an important role in the 
reionization of the Universe at $z$ $>$ 5 \citep{O09}.

\section*{Acknowledgements}

We thank D. M. Terndrup and C. Wiens for help with the LBT observations.
We are grateful to anonymous referees for useful comments on the manuscript.
Funding for the Sloan Digital Sky Survey IV has been provided by
the Alfred P. Sloan Foundation, the U.S. Department of Energy Office of
Science, and the Participating Institutions. SDSS-IV acknowledges
support and resources from the Center for High-Performance Computing at
the University of Utah. The SDSS web site is www.sdss.org.
SDSS-IV is managed by the Astrophysical Research Consortium for the 
Participating Institutions of the SDSS Collaboration. 
This research has made use of the NASA/IPAC Extragalactic Database (NED), which 
is operated by the Jet Propulsion Laboratory, California Institute of 
Technology, under contract with the National Aeronautics and Space 
Administration.

\input{ref.tex}
\bsp

\label{lastpage}

\end{document}

%% file: tab1_1.tex
\begin{table}
\caption{Observed characteristics of J0811$+$4730 \label{tab1}}
\begin{tabular}{lr} \hline
Parameter                 &  Value       \\ \hline
R.A.(J2000)               &  08:11:52.12 \\
Dec.(J2000)               & +47:30:26.24 \\
  $z$                     &  0.04444$\pm$0.00003     \\
  $g$, mag                &   21.37$\pm$0.05      \\
  $M_g$, mag$^\dag$        & $-$15.41$\pm$0.06     \\
log $M_\star$/M$_\odot$$^\ddag$&   6.24$\pm$0.33       \\
log $M_\star$/M$_\odot$$^{\ddag\ddag}$&   6.29$\pm$0.06       \\
$L$(H$\beta$), erg s$^{-1}$$^*$&(2.1$\pm$0.1)$\times$10$^{40}$\\
SFR, M$_\odot$yr$^{-1}$$^{\dag\dag}$  &     0.48$\pm$0.02 \\
\hline
  \end{tabular}

\noindent$^\dag$Corrected for Milky Way extinction.

\noindent$^\ddag$Derived from the extinction- and aperture-corrected SDSS 
spectrum.

\noindent$^{\ddag\ddag}$Derived from the extinction- and aperture-corrected LBT 
spectrum.

\noindent$^*$Corrected for extinction and the SDSS spectroscopic aperture.

\noindent$^{\dag\dag}$Derived from the \citet{K98} relation using the extinction- 
and aperture-corrected H$\beta$ luminosity.
  \end{table}

%% file: tab2.tex
\begin{table}
\caption{Observed and extinction-corrected emission-line fluxes \label{tab2}}
\begin{tabular}{lrr} \hline
Line&100$\times$$F$($\lambda$)/$F$(H$\beta$)&100$\times$$I$($\lambda$)/$I$(H$\beta$) \\ \hline
3187.74 He {\sc i}              &  2.71$\pm$0.66&  3.32$\pm$0.81\\
3203.10 He {\sc ii}             &  1.71$\pm$0.89&  2.09$\pm$1.02\\
3703.30 H16                     &  1.35$\pm$0.14&  3.41$\pm$0.36\\
3711.97 H15                     &  1.83$\pm$0.20&  3.72$\pm$0.49\\
3721.94 H14                     &  2.22$\pm$0.18&  5.83$\pm$0.90\\
3727.00 [O {\sc ii}]            & 14.70$\pm$0.33& 16.75$\pm$0.40\\
3734.37 H13                     &  1.52$\pm$0.23&  3.61$\pm$1.00\\
3750.15 H12                     &  2.73$\pm$0.23&  4.73$\pm$0.45\\
3770.63 H11                     &  2.85$\pm$0.19&  4.89$\pm$0.40\\
3797.90 H10                     &  4.48$\pm$0.19&  6.66$\pm$0.36\\
3819.64 He {\sc i}              &  1.18$\pm$0.16&  1.33$\pm$0.18\\
3835.39 H9                      &  6.37$\pm$0.22&  8.86$\pm$0.38\\
3868.76 [Ne {\sc iii}]          & 11.91$\pm$0.28& 13.39$\pm$0.32\\
3889.00 He {\sc i}+H8           & 17.26$\pm$0.34& 20.96$\pm$0.48\\
3968.00 [Ne {\sc iii}]+H7       & 18.05$\pm$0.34& 21.69$\pm$0.47\\
4026.19 He {\sc i}              &  1.59$\pm$0.14&  1.76$\pm$0.15\\
4068.60 [S {\sc ii}]            &  0.29$\pm$0.09&  0.31$\pm$0.10\\
4101.74 H$\delta$               & 23.26$\pm$0.40& 26.83$\pm$0.51\\
4120.84 He {\sc i}              &  0.39$\pm$0.09&  0.43$\pm$0.10\\
4340.47 H$\gamma$               & 44.14$\pm$0.69& 48.13$\pm$0.80\\
4363.21 [O {\sc iii}]           &  5.93$\pm$0.19&  6.26$\pm$0.20\\
4387.93 He {\sc i}              &  0.18$\pm$0.09&  0.18$\pm$0.09\\
4471.48 He {\sc i}              &  3.51$\pm$0.15&  3.65$\pm$0.16\\
4658.10 [Fe {\sc iii}]          &  0.51$\pm$0.10&  0.52$\pm$0.10\\
4685.94 He {\sc ii}             &  2.24$\pm$0.15&  2.27$\pm$0.15\\
4712.00 [Ar {\sc iv}]+He {\sc i}&  0.91$\pm$0.16&  0.91$\pm$0.16\\
4740.20 [Ar {\sc iv}]           &  0.57$\pm$0.11&  0.58$\pm$0.11\\
4861.33 H$\beta$                &100.00$\pm$1.50&100.00$\pm$1.52\\
4921.93 He {\sc i}              &  0.97$\pm$0.13&  0.95$\pm$0.13\\
4958.92 [O {\sc iii}]           & 56.75$\pm$0.88& 55.60$\pm$0.87\\
4988.00 [Fe {\sc iii}]          &  0.44$\pm$0.09&  0.43$\pm$0.09\\
5006.80 [O {\sc iii}]           &165.79$\pm$2.46&161.59$\pm$2.42\\
5015.68 He {\sc i}              &  1.91$\pm$0.13&  1.86$\pm$0.13\\
5875.60 He {\sc i}              & 10.38$\pm$0.18&  9.45$\pm$0.17\\
6300.30 [O {\sc i}]             &  0.65$\pm$0.06&  0.60$\pm$0.05\\
6312.10 [S {\sc iii}]           &  0.46$\pm$0.06&  0.41$\pm$0.06\\
6548.10 [N {\sc ii}]            &  0.38$\pm$0.02&  0.33$\pm$0.02\\
6562.80 H$\alpha$               &311.93$\pm$4.56&273.87$\pm$4.38\\
6583.40 [N {\sc ii}]            &  0.72$\pm$0.06&  0.63$\pm$0.06\\
6678.10 He {\sc i}              &  2.97$\pm$0.08&  2.59$\pm$0.08\\
6716.40 [S {\sc ii}]            &  1.39$\pm$0.06&  1.21$\pm$0.05\\
6730.80 [S {\sc ii}]            &  1.23$\pm$0.06&  1.07$\pm$0.05\\
7065.30 He {\sc i}              &  3.88$\pm$0.09&  3.31$\pm$0.08\\
7135.80 [Ar {\sc iii}]          &  0.99$\pm$0.06&  0.84$\pm$0.05\\
8545.00 P15                     &  1.31$\pm$0.10&  1.23$\pm$0.10\\
8665.00 P13                     &  0.99$\pm$0.10&  0.95$\pm$0.10\\
8750.00 P12                     &  1.23$\pm$0.10&  1.13$\pm$0.10\\
8863.00 P11                     &  1.70$\pm$0.13&  1.48$\pm$0.11\\
9015.00 P10                     &  2.43$\pm$0.15&  2.06$\pm$0.13\\
9069.00 [S {\sc iii}]           &  2.14$\pm$0.18&  1.81$\pm$0.15\\
9229.00 P9                      &  2.94$\pm$0.20&  2.45$\pm$0.17\\
$C$(H$\beta$)$^{\rm a}$         &\multicolumn{2}{c}{0.165$\pm$0.019}\\
$F$(H$\beta$)$^{\rm b}$         &\multicolumn{2}{c}{12.60$\pm$0.05}\\
EW(H$\beta$)$^{\rm c}$          &\multicolumn{2}{c}{282.0$\pm$1.0}\\
EW(abs)$^{\rm c}$               &\multicolumn{2}{c}{2.8$\pm$0.3}\\
\hline
  \end{tabular}

\hbox{$^{\rm a}$Extinction coefficient, derived from the observed hydrogen} 

\hbox{\,~Balmer decrement.}

\hbox{$^{\rm b}$Observed flux in units of 10$^{-16}$ erg s$^{-1}$ cm$^{-2}$.}

\hbox{$^{\rm c}$Equivalent width in \AA.}

  \end{table}

%% file: tab3_2.tex
\begin{table}
\caption{Electron temperatures and element abundances \label{tab3}}
\begin{tabular}{lccccc} \hline
Property                             &Value          \\ \hline
$T_{\rm e}$(O {\sc iii}), K          &21700$\pm$500       \\
$T_{\rm e}$(O {\sc ii}), K           &15600$\pm$500       \\
$T_{\rm e}$(S {\sc iii}), K          &19700$\pm$500       \\
$N_{\rm e}$(S {\sc ii}), cm$^{-3}$    &380$\pm$140         \\ \\
O$^+$/H$^+$$\times$10$^6$            &1.405$\pm$0.115 \\
O$^{2+}$/H$^+$$\times$10$^6$          &7.884$\pm$0.398 \\
O$^{3+}$/H$^+$$\times$10$^6$          &0.244$\pm$0.021 \\
O/H$\times$10$^6$                   &9.532$\pm$0.415 \\
12+log(O/H)                         &6.979$\pm$0.019     \\ \\
N$^+$/H$^+$$\times$10$^7$            &0.404$\pm$0.053 \\
ICF(N)                              &6.384 \\
N/H$\times$10$^7$                   &2.785$\pm$0.358 \\
log(N/O)                            &$-$1.535$\pm$0.044~~~\\ \\
Ne$^{2+}$/H$^+$$\times$10$^6$        &1.409$\pm$0.072 \\
ICF(Ne)                             &1.073 \\
Ne/H$\times$10$^6$                  &1.512$\pm$0.087 \\
log(Ne/O)                           &$-$0.800$\pm$0.031~~~\\ \\
S$^+$/H$^+$$\times$10$^7$            &0.213$\pm$0.021 \\
S$^{2+}$/H$^+$$\times$10$^7$         &0.996$\pm$0.146 \\
ICF(S)                              &1.621 \\
S/H$\times$10$^7$                   &1.961$\pm$0.237 \\
log(S/O)                            &$-$1.687$\pm$0.056~~~\\ \\
Ar$^{2+}$/H$^+$$\times$10$^8$        &2.174$\pm$0.142 \\
Ar$^{3+}$/H$^+$$\times$10$^8$        &2.482$\pm$0.494 \\
ICF(Ar)                             &1.223 \\
Ar/H$\times$10$^8$                  &2.659$\pm$0.628 \\
log(Ar/O)                           &$-$2.555$\pm$0.104~~~\\ \\
Fe$^{2+}$/H$^+$$\times$10$^7$        &0.898$\pm$0.185 \\
ICF(Fe)                             &9.263 \\
Fe/H$\times$10$^7$                  &8.315$\pm$1.716 \\
log(Fe/O)                           &$-$1.060$\pm$0.092~~~\\
\hline
  \end{tabular}
  \end{table}

%% file: tab4.tex
\begin{table*}
\caption{Parameters of the lowest-metallicity SFGs \label{tab4}}
\begin{tabular}{lcrcrrccr} \hline
Name                             &12+logO/H&$M_B$&$M_g$&O$_3$&O$_{32}$&R$_{23}$&N$_{2}$&Ref.          \\ \hline
J0811$+$4730        &6.98$\pm$0.02& ...~~ &$-$15.4&1.61& 9.65&2.34&0.0023&1\\
SBS0335$-$052W\#2   &7.01$\pm$0.07&$-$14.7& ...   &1.48& 4.91&2.27& ...  &2\\
AGC198691           &7.02$\pm$0.03& ...~~ &$-$11.5&1.28& 2.74&2.17&0.0058&3\\
SBS0335$-$052E\#7   &7.12$\pm$0.04&$-$17.3& ...   &1.94& 7.78&2.84&0.0031&2\\
IZw18NW             &7.16$\pm$0.01& ...~~ &$-$15.3&1.95& 6.83&2.89&0.0034&4\\
LeoP                &7.17$\pm$0.04&$-$8.8 & ...   &1.45& 3.12&2.40&0.0091&5\\
J1220$+$4915        &7.18$\pm$0.03& ...~~ &$-$12.8&2.97&13.48&4.19&0.0051&6\\
IZw18SE             &7.19$\pm$0.02& ...~~ &$-$15.3&1.60& 2.68&2.73&0.0072&4\\
DDO68               &7.20$\pm$0.05&$-$15.8& ...   &1.89& 3.26&2.47&0.0057&7\\
SBS0335$-$052W\#1   &7.22$\pm$0.07&$-$14.7& ...   &1.30& 1.73&2.49&0.0122&2\\
J2104$-$0035        &7.24$\pm$0.02& ...~~ &$-$12.2&2.79&13.95&3.92&0.0025&6\\
SBS0335$-$052E\#1+2 &7.28$\pm$0.01&$-$17.3& ...   &3.03&13.90&4.27&0.0036&2\\
\hline
  \end{tabular}

{\bf Notes.} O$_3$=[O {\sc iii}]$\lambda$5007/H$\beta$, O$_{32}$=[O {\sc iii}]$\lambda$5007/[O {\sc ii}]$\lambda$3727, R$_{23}$=([O {\sc ii}]$\lambda$3727+[O {\sc iii}]$\lambda$4959+[O {\sc iii}]$\lambda$5007)/H$\beta$, N$_2$=[N~{\sc ii}]$\lambda$6584/H$\alpha$.~~~~\\

{\bf References.} 
1 - this paper, 2 - \citet{I09}, 3 - \citet{H16}, 4 - \citet{IT98}, 
5 - \citet{S13}, \\ 
6 - Izotov et al., in preparation, 7 - \citet{B12}.~~~~~~~~~~~~~~~~~~~~~~~~~~~~~~~~~~~~~~~~~~~~~~~~~~~~~~~~~~~~~~~~~~~~~~~~~~~~~~~~~~~~~~~~~~~~~~~~~~~~~~~~~~~~~~~
  \end{table*}

%% file: text14-clean.bbl
\begin{thebibliography}{}



\bibitem[Albareti et al.(2016)]{A16} Albareti F. D. et al., 2016, \apjs, in press; preprint arXiv:1608.02013

\bibitem[Aller(1984)]{A84} Aller L. H., 1984, Physics of Thermal
Gaseous Nebulae (Dordrecht: Reidel)

\bibitem[Aloisi et al.(2007)]{A07} Aloisi A. et al., 2007, \apj, 667, L151


\bibitem[Annibali et al.(2013)]{A13} Annibali F. et al., 2013, \aj, 146, 144

\bibitem[Asari et al.(2007)]{As07} Asari N. V., Cid Fernandes R., 
Stasi\'nska G., Torres-Papaqui J. P., Mateus A., Sodr\'e L. Jr., 
Schoenell W., Gomes J. M., 2007, \mnras, 381, 263

\bibitem[Baldwin et al.(1981)Baldwin, Phillips \& Terlevich]{BPT81} 
Baldwin J. A., Phillips M. M., Terlevich R., 1981, \pasp, 93, 5




\bibitem[Berg et al.(2012)]{B12} Berg D. A. et al., 2012, \apj, 754, 98

\bibitem[Binney \& Merrifield(1998)]{BM98} Binney J., Merrifield M., 1998, 
Galactic Astronomy (Princeton University Press)

\bibitem[Blanton et al.(2003)]{B03} Blanton M. R. et al., 2003, \apj, 592, 819 











\bibitem[Cardamone et al.(2009)]{Ca09} Cardamone C. et al.,
2009, \mnras, 399, 1191

\bibitem[Cardelli et al.(1989)Cardelli, Clayton \& Mathis]{C89} 
Cardelli J. A., Clayton G. C., Mathis J. S., 1989, \apj, 345, 245

\bibitem[Ch\'avez et al.(2012)]{C12} Ch\'avez R., Terlevich E., Terlevich R.,
Plionis M., Bresolin F., Basilakos S., Melnick J., 2012, \mnras, 425, L56

\bibitem[Cid Fernandes et al.(2005)]{CF05} Cid Fernandes R., Mateus A.,
Sodr\'e L. Jr., Stasi\'nska G., Gomes J. M., 2005, \mnras, 358, 363

\bibitem[Ekta \& Chengalur(2010)]{EC10} Ekta B, Chengalur J. N., 2010,
\mnras, 406, 1238





\bibitem[Ferland et al.(1998)]{F98} Ferland G. J., Korista K. T.,
Verner D. A., Ferguson J. W., Kingdon J. B., Verner E. M., 1998,
\pasp, 110, 761

\bibitem[Ferland et al.(2013)]{F13} Ferland G. J. et al., 2013, \rmxaa, 
49, 137

\bibitem[Filippenko(1982)]{F82} Filippenko A. V., 1982, \pasp, 
94, 715





\bibitem[Gallazzi \& Bell(2009)]{GB09} Gallazzi A., Bell E. F., 2009, \apjs,
185, 253 


\bibitem[Giovanelli et al.(2005)]{G05} Giovanelli R. et al., 2005, \aj, 130, 2598

\bibitem[Girardi et al.(2000)]{G00} Girardi L., Bressan A., Bertelli G., 
Chiosi C., 2000, \aaps, 141, 371




\bibitem[Guseva et al.(2017)]{G17} Guseva N. G., Izotov Y. I., Fricke K. J.,
Henkel C., 2017, \aap, 599, A65

\bibitem[Haynes et al.(2011)]{H11} Haynes M. P. et al., 2011, \aj, 142, 170

\bibitem[Hirschauer et al.(2016)]{H16} Hirschauer A. S. et al., 2016, \apj, 822,
108




\bibitem[Izotov \& Thuan(1998)]{IT98} Izotov Y. I., Thuan T. X., 
1998, \apj, 497, 227



\bibitem[Izotov et al.(1994)Izotov, Thuan \& Lipovetsky]{ITL94} Izotov Y. I.,
Thuan T. X., Lipovetsky V. A., 1994, \apj, 435, 647


\bibitem[Izotov et al.(2005)Izotov, Thuan \& Guseva]{I05} Izotov Y. I., 
Thuan T. X., Guseva N. G., 2005, \apj, 632, 210

\bibitem[Izotov et al.(2006)]{I06} Izotov Y. I., Stasi\'nska G., Meynet G., 
Guseva N. G., Thuan T. X., 2006, \aap, 448, 955



\bibitem[Izotov et al.(2009)]{I09} Izotov Y. I., Guseva N. G., Fricke K. J.,
Papaderos P., 2009, \aap, 503, 61

\bibitem[Izotov et al.(2011)Izotov, Guseva \& Thuan]{I11} Izotov Y. I., 
Guseva N. G., Thuan T. X., 2011, \apj, 728, 161

\bibitem[Izotov et al.(2012)Izotov, Thuan \& Guseva]{I12} Izotov Y. I., 
Thuan T. X., Guseva N. G., 2012, \aap, 546, 122



\bibitem[Izotov et al.(2014)]{I14} Izotov Y. I., Guseva N. G., 
Fricke K. J., Kr\"ugel E., Henkel C., 2014, \aap, 570, A97


\bibitem[Izotov et al.(2015)]{I15} Izotov Y. I., Guseva N. G., 
Fricke K. J., Henkel C., 2015, \mnras, 451, 2251



\bibitem[Izotov et al.(2016)]{I16c} Izotov Y. I., Guseva N. G.,
Fricke K. J., Henkel C., 2016, \mnras, 462, 4427

\bibitem[Izotov et al.(2017)Izotov, Thuan \& Guseva]{I17} Izotov Y. I., 
Thuan T. X., Guseva N. G., 2017, \mnras, 471, 548


\bibitem[Karman et al.(2017)]{K17} Karman W. et al., 2016, \aap, 599, A28


\bibitem[Kauffmann et al.(2003)]{K03} Kauffmann G. et al., 2003, \mnras, 341, 33

\bibitem[Kennicutt(1998)]{K98} Kennicutt R. C., Jr.,
1998, \araa, 36, 189




\bibitem[Law et al.(2011)]{L11} Law D. R. et al., 2011, \aj, 152, 83




\bibitem[Leitherer et al.(1999)]{L99} Leitherer C. et al., 1999, \apjs, 123, 3

\bibitem[Leitherer et al.(2014)]{L14} Leitherer C., Ekstr\"om S., 
Meynet G., Schaerer D., Agienko K. B., Levesque E. M., 2014, \apjs, 212, 14


\bibitem[Lejeune et al.(1997)Lejeune, Buser \& Cuisiner]{L97} 
Lejeune T., Buser R., Cuisinier F., 1997, \aaps, 125, 229













\bibitem[Ouchi et al.(2009)]{O09} Ouchi M. et al., 2009, \apj, 706, 1136


\bibitem[Pettini \& Pagel(2004)]{PP04} Pettini M., Pagel B. E. J., 2004,
\mnras, 348, L59


\bibitem[Pilyugin \& Thuan(2005)]{PT05} Pilyugin L., Thuan T.X., 2005,
\apj, 631, 231

\bibitem[Planck Collaboration XVI(2014)]{P14} Planck Collaboration XVI,
2014, \aap, 571, A16





\bibitem[Pustilnik et al.(2004)Pustilnik, Pramskij \& Kniazev]{P04} 	
Pustilnik S. A., Pramskij A. G., Kniazev A. Y., 2004, \aap, 425, 51 

\bibitem[Pustilnik et al.(2005)Pustilnik, Kniazev \& Pramskij]{P05} 	
Pustilnik S. A., Kniazev A. Y., Pramskij A. G., 2005, \aap, 443, 91








\bibitem[Rudolf et al.(2016)]{R16} Rudolf N., G\"unther H. M., Schneider P.C., and Schmitt J. H. M. M., 2016, \aap, 585, 113

\bibitem[Salpeter(1955)]{S55} Salpeter E. E., 1955, \apj, 121, 161




\bibitem[Searle \& Sargent(1972)]{SS72} Searle L., Sargent W. L. W., 1972,
\apj, 173, 25


\bibitem[Skillman \& Kennicutt(1993)]{SK93} Skillman E., Kennicutt R. C. Jr.,
 1993, \apj, 411, 655

\bibitem[Skillman et al.(2013)]{S13} Skillman E. D. et al., 2013, \aj, 146, 3


\bibitem[Stasi\'nska et al.(2015)]{S15} Stasi\'nska G., Izotov Y., 
Morisset C.,  Guseva N., 2015, \aap, 576, A83






\bibitem[Thuan et al.(1997)Thuan, Izotov \& Lipovetsky]{TIL97} Thuan T. X., 
Izotov Y. I., Lipovetsky V. A., 1997, \apj, 477, 661


\bibitem[Tremonti et al.(2004)]{T04} Tremonti C. et al., 2004, \apj, 613, 898








\bibitem[Wright(2006)]{W06} Wright E. L., 2006, \pasp, 118, 1711








\end{thebibliography}
